\documentclass[manuscript]{acmart}
\AtBeginDocument{%
  \providecommand\BibTeX{{%
    \normalfont B\kern-0.5em{\scshape i\kern-0.25em b}\kern-0.8em\TeX}}}

\acmJournal{CSUR}
\usepackage{xspace}	% Custom
\usepackage{cleveref}
\usepackage{booktabs}
\usepackage{todonotes}
\usepackage{longtable}
\usepackage{array}
\usepackage{multirow}
\usepackage{supertabular}
\usepackage{tcolorbox}
\tcbuselibrary{skins}
\usetikzlibrary{calc}
\usepackage{tikz}
\usepackage{lscape}
\usepackage{xparse}

\usepackage[inline]{enumitem}

\def\bb{block-based\xspace}
\def\Bb{Block-based\xspace}
\def\bbe{\bb environment}
\def\bbes{\bb environments\xspace}
\def\Bbe{\Bb environment}
\def\Bbes{\Bb environments\xspace}
\NewDocumentCommand{\rot}{O{45} O{1em} m}{\makebox[#2][l]{\rotatebox{#1}{#3}}}
\begin{document}
\newcounter{mycounter}

\newcommand\sbullet[1][.5]{\mathbin{\vcenter{\hbox{\scalebox{1.75}{$\bullet$}}}}}

%\tcbset{tool/.style={enhanced,
%enlarge top by=10mm,
%overlay={%
%\path[fill=blue!65,line width=.4mm] (frame.north west)--++(17mm,0)coordinate(n2)--++(0,8mm)--++(-20mm,0) arc (-90:90:-4mm)--cycle;
%\node at ([shift={(5mm,4mm)}]frame.north west){\color{white}{\textbf{\sffamily Language}}};
%\path[fill=blue!65] ([xshift=.4mm]n2)--++(0,8mm)--++(7mm,0)--++(0,-8mm)--cycle;
%\node at ([shift={(4mm,4mm)}]n2){\color{white}{\textbf{\sffamily \themycounter}}};
%\node at ([shift={(18mm,4mm)}]n2){\itshape\textbf{\sffamily Language}};
%}}}

\tcbset{mystyle/.style={
  enhanced,
  outer arc=0pt,
  arc=0pt,
  attach boxed title to top left,
  fonttitle=\sffamily
  }
}

\newtcolorbox[auto counter]{mybox2}[1][htb]{
  mystyle,
  title=Category~\thetcbcounter,
  overlay unbroken and first={
      \path
        let
        \p1=(title.north east),
        \p2=(frame.north east)
        in
        node[anchor=west,font=\sffamily,text width=\x2-\x1] 
        at (title.east) {#1};
  }
}

%% BETTER FLOATS
\renewcommand{\topfraction}{0.9}	% max fraction of floats at top
\renewcommand{\bottomfraction}{0.8}	% max fraction of floats at bottom
    %   Parameters for TEXT pages (not float pages):
\setcounter{topnumber}{2}
\setcounter{bottomnumber}{2}
\setcounter{totalnumber}{4}     % 2 may work better
\setcounter{dbltopnumber}{2}    % for 2-column pages
\renewcommand{\dbltopfraction}{0.9}	% fit big float above 2-col. text
\renewcommand{\textfraction}{0.07}	% allow minimal text w. figs
%   Parameters for FLOAT pages (not text pages):
\renewcommand{\floatpagefraction}{0.7}	% require fuller float pages
% N.B.: floatpagefraction MUST be less than topfraction !!
\renewcommand{\dblfloatpagefraction}{0.7}	% require fuller float pages
%% END OF BETTER FLOATS
    
% Blue box

%\tcbset{mystyle/.style={
%  enhanced,
%  enlarge top by=10mm,
%  outer arc=0pt,
%  arc=0pt,
%  attach boxed title to top left,
%  fonttitle=\sffamily
%  }
%}

%\newtcolorbox[auto counter]{mybox}[2][htb]{
%  mystyle,
%%  title=Example~\thetcbcounter,
%  overlay={
%  \path[fill=blue!65,line width=.4mm] (frame.north west)--++(17mm,0)coordinate(n2)--++(0,8mm)--++(-20mm,0) arc (-90:90:-4mm)--cycle;
%\node at ([shift={(5mm,4mm)}]frame.north west){\color{white}{\textbf{\sffamily {#1}}}};
%\path[fill=blue!65] ([xshift=.4mm]n2)--++(0,8mm)--++(7mm,0)--++(0,-8mm)--cycle;
%\node at ([shift={(4mm,4mm)}]n2){\color{white}{\textbf{\sffamily \thetcbcounter}}};
%\node at ([shift={(30mm,4mm)}]n2){\textbf{\sffamily {#2}}};}
%}

%\title{Mapping Study on Visual Block-based Languages and Tools}
\title{DRAFT-What you always wanted to know but could not find about block-based environments}

\author{Mauricio Verano Merino}
\authornotemark[1]
\email{m.verano.merino@tue.nl}
\affiliation{%
  \institution{Eindhoven University of Technology - CWI}
%  \streetaddress{P.O. Box 1212}
  \city{Eindhoven}
  \country{The Netherlands}
}

\author{Jurgen Vinju}
\affiliation{%
  \institution{CWI - Eindhoven University of Technology}
%  \streetaddress{1 Th{\o}rv{\"a}ld Circle}
  \city{Amsterdam}
  \country{The Netherlands}}
\email{Jurgen.Vinju@cwi.nl}

\author{Mark van den Brand}
\affiliation{%
  \institution{Eindhoven University of Technology}
%  \streetaddress{1 Th{\o}rv{\"a}ld Circle}
  \city{Eindhoven}
  \country{The Netherlands}}
\email{M.G.J.v.d.Brand@tue.nl}

%\affiliation{%
%  \institution{Inria Paris-Rocquencourt}

\renewcommand{\shortauthors}{Verano Merino, et al.}

\begin{abstract}
\Bbes are visual programming environments, which are becoming more and more popular because of their ease of use.
The ease of use comes thanks to their intuitive graphical representation and structural metaphors (jigsaw-like puzzles) to display valid combinations of language constructs to the users.
Part of the current popularity of \bbes is thanks to Scratch. As a result they are often associated with tools for children or young learners.
However, it is unclear how these types of programming environments are developed and used in general.
So we conducted a systematic literature review on \bbes by studying 152 papers published between 2014 and 2020, and a non-systematic tool review of 32 \bbes.
%to provide a landscape of the field.
In particular, we provide a helpful inventory of \bb editors for end-users on different topics and domains. 
Likewise, we focused on identifying the main components of \bbes, how they are engineered, and how they are used.
This survey should be equally helpful for language engineering researchers and language engineers alike.
% TODO change order
\end{abstract}

\begin{CCSXML}
<ccs2012>
   <concept>
       <concept_id>10002944.10011122.10002945</concept_id>
       <concept_desc>General and reference~Surveys and overviews</concept_desc>
       <concept_significance>500</concept_significance>
       </concept>
   <concept>
       <concept_id>10011007.10011006.10011050.10011052</concept_id>
       <concept_desc>Software and its engineering~Graphical user interface languages</concept_desc>
       <concept_significance>500</concept_significance>
       </concept>
   <concept>
       <concept_id>10011007.10011006.10011050.10011058</concept_id>
       <concept_desc>Software and its engineering~Visual languages</concept_desc>
       <concept_significance>500</concept_significance>
       </concept>
   <concept>
       <concept_id>10011007.10011006.10011066.10011069</concept_id>
       <concept_desc>Software and its engineering~Integrated and visual development environments</concept_desc>
       <concept_significance>300</concept_significance>
       </concept>
 </ccs2012>
\end{CCSXML}

\ccsdesc[500]{General and reference~Surveys and overviews}
\ccsdesc[500]{Software and its engineering~Graphical user interface languages}
\ccsdesc[500]{Software and its engineering~Visual languages}
\ccsdesc[300]{Software and its engineering~Integrated and visual development environments}

\keywords{block-based environments, visual languages, programming environments, language engineering}
% TODO keywords

\maketitle

\section{Introduction}
\begin{figure}[htp]
  \includegraphics[width=0.1\textwidth]{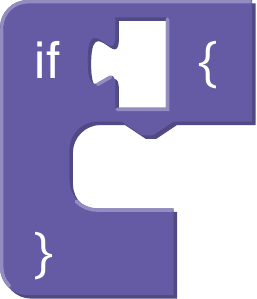}
  \caption{\Bb representation of an \texttt{if} statement.}
  \label{fig:if}
\end{figure}

In the past decade, end-user programming environments have become more popular and more relevant. End-users significantly outnumber professional programmers~\cite{Rough:2020}.
These environments allow end-users to create and adapt software, enabling them to achieve a myriad of tasks.
Without them, most of these tasks would not be possible unless one has a background in computer science or software engineering.
\Bbes are part of this set of end-user visual programming environments.
One of their main characteristics is the editor, which presents the language constructs to the end-user as  graphical elements that resemble Lego blocks.
Each of these blocks is chararacterized by visual signifiers (e.g., color and shape) that hint end-users of its semantics and its (possible) connections to other blocks.
An example of a \bb representation of an \texttt{if} statement is shown in \Cref{fig:if}.
The benefit of having a \bb editor is to offer a programming experience based on what-you-see-is-what-you-get (WYSIWYG) and the impossibility of syntactic errors~\cite{Price:2015,Moors:2017,Weintrop:20172,Weintrop:20184}.
Moreover, these editors support different \bb programming paradigms, such as \textit{configuration}, \textit{serial}, \textit{parallel}, and \textit{event-driven}~\cite{bb-paradigms}.

The popularity of \bb editors have increased in recent years, partially due to Scratch's popularity (23rd most popular programming language~\cite{tiobe:2021}).
However, languages that provide such a type of editors are not new, yet \bb editors have been mainly used and associated with computer science education or applications for children.
This paper explores whether this is true or not.
In case this claim is not true, we explore how these programming environments have been adopted beyond the realms of education or children.
Moreover, this paper identifies \bbes' main components to understand them and increase their adoption in different domains and for different user groups; and studies whether the development of \bbes is supported by specialized language engineering tooling (e.g., language workbenches).

To have a clear overview of the landscape of \bb environments and understand how they are developed, we conducted a systematic and a less-systematic literature review.
A systematic literature review collects and summarizes all the existing research evidence of a domain and identifies possible gaps in current research~\cite{kitchenham2007guidelines}.
Initially we started with the less-systematic method, in which we sought \bbes and their features. We ran into the limits of this ad-hoc method and continued with an SLR (structured literature review) to identify possible gaps in current research~\cite{kitchenham2007guidelines}.

Since there exists no primary conference or journal focused on block-based environments, we expect that papers on this topic are spread over different academic communities with different characteristics. The papers we found in the venues will frame the answers to the research questions about \bbes.% and so we ask the first question about the publications themselves in order to clearly document that frame.

%The results of the non-systematic review are presented in \Cref{sec:non-syst}.

The contributions of this paper are summarized as follows:
\begin{itemize}
	\item A systematic literature review on \bbes\, which provides an overview of the main features of \bbes, the landscape in which these programming environments are used, publication venues, programming languages used in their development, and the most popular environments (\Cref{sec:systematic}).
	\item A deeper (qualitative) understanding of \bbes\ and their components (\Cref{sec:qone}).
	\item An understanding of how \bbes\ are implemented and the tools and languages involved in their development (\Cref{sec:qthree,ssec:qfour}).
	\item A non-systematic tool review of \bbes that presents some of the most relevant features of these programming environments (\Cref{sec:non-syst}).
\end{itemize}
We describe the threats to the validity of this study in \Cref{sec:threats} and a discussion of the results obtained in this study in \Cref{sec:disc}. The non-systematic literature review fills some gaps of the systematic literature review and vice versa. We concludes with a discussion of related work, conclusions and future research directions (\Cref{sec:related,sec:conclusions}).

\section{Systematic Review: Motivation and Methodology}
\label{sec:met}

A literature review about \bbes can be approached from different perspectives. Different perspectives are relevant because \bbes are used across different domains and user groups, and developed by engineers with different backgrounds as well. Our agenda is to make \bb editors part of the default set of services offered by language workbenches~\cite{Fowler} and to investigate how editor generation technology (metaprogramming) is used to support this. Therefore, our primary motivation in this literature review is to understand in which contexts are \bbes used and how exactly they are developed.

We aim to make \bbes\ available for all DSLs developed using a language workbench (LWB).
LWBs have shown benefits by offering specialized (generative) technology for developing software languages and their tooling. LWBs are used to define both the syntax and semantics of languages. Based on these two, language engineers can also semi-automatically derive a full-featured IDE for their language~\cite{meta-environment,imp,stratego-xt}. We believe that \bb editors can also be derived from existing language components. To enable that research we need a clear set of requirements to frame the contributions of a block-based LWB and we must understand how \bbes are best implemented. The current SLR helps in achieving these goals.

\subsection{Need for a systematic review}
Coronado et al.~\cite{CORONADO:2020} published a literature review about using visual programming environments for programming robots.
In their work, they present an overview of the environments that use a visual editor for programming robots; this considers all sorts of graphical elements to represent language constructs, including \bb editors.
However, their main target was to understand the tools used to program robots.
A similar approach can be used for other purposes, for instance, to identify the programming environments used for teaching computational skills.
Our view is that it is essential to understand \bbes in general; this includes understanding their features, applications, and technologies involved in their development.

We aim to increase the adoption of the \bb metaphor beyond the realm of computing education, make its development part of the set of generic services offered by specialized tooling for creating software languages (i.e., language workbenches), and explore further applications in industrial settings.

\subsection{Protocol}
% Protocol
This section presents the protocol defined for the systematic review.
We followed Kitchenham et al.~\cite{kitchenham2007guidelines,kitchenham:2009} guidelines for conducting a systematic literature review in software engineering.
The primary goal is to summarize existing evidence in literature of the development and usage of \bbes, in different disciplines.
The secondary goal is to identify possible gaps between the development of these visual programming environments and existing tools and technologies specialized in developing software languages and their tooling.

\subsubsection{Scope}
Visual programming environments are environments that rely on graphical editors for building programs.
Different notations are adopted by these environments, such as flow charts, UML diagrams, and \bb editors.
In this survey, we focus our attention on the latter notation.

\Bbes are not new, but the popularity of Scratch\footnote{\url{https://scratch.mit.edu/}} has inspired a new wave of visual programming environments.
These environments are characterized by representing language constructs with graphical blocks that resemble Lego blocks.
Moreover, these environments offer visual cues that help users understand what are the possibilities for connecting blocks.

\subsubsection{Research questions}\label{ssec:research}
The research questions addressed in this study are:

\def\qzero{\textbf{RQ0}\xspace}
\def\bzero{\textit{What are the characteristics of the papers that present \bb editors?}}
\def\qone{\textbf{RQ1}\xspace}
\def\bpone{\textit{What are the components of a \bbe?}}
\def\qtwo{\textbf{RQ2}\xspace}
\def\bptwo{\textit{What are the tools used to develop \bbes?}}
\def\qthree{\textbf{RQ3}\xspace}
\def\bpthree{\textit{How are \bbes\ developed?}}
\def\qfour{\textbf{RQ4}\xspace}
\def\bpfour{\textit{What languages offer a \bb editor and what are these languages used for?}}
%What languages offer a built-in \bb editor, and where are these languages being used for?
%\def\bpfour{\textit{Where are \bb environments being used?}}
%How are \bb environments being implemented?

\begin{quote}
\qzero \bzero

\qone \bpone

\qtwo \bptwo

\qthree \bpthree

\qfour \bpfour\\
\end{quote}

The motivation for the meta question \qzero is that we expect publications on block-based editors to be scattered acros many different (types of) venues: from fundamental computer science all the way to applications in other academic domains such as medicine, and everything in between. The answer to \qzero helps to frame the answers to the following research questions. Research questions \qzero, \qtwo, and \qthree are answered through the systematic review.
\qone and \qfour are answered using both the systematic and the non-systematic approach.

\subsection{Search process}

Languages that use a \bb editor are becoming popular outside the academic world for their ease of use. 
For instance, commercial robots, programmable microcontrollers, and applications for children use them as an effective end-user interface. 
Consequently, many of these languages have been developed outside the academic world, which means that there are language implementations that do not have a corresponding academic publication. Vice versa there exist academic publications about languages which do not have an implementation (anymore).

Therefore, to obtain a complete overview of the landscape, it is essential to include both academic and non-academic tools in this literature review.
Therefore, we decided to follow a combined search process that is both fully systematic and less-systematic. 
%As a result, to include non-academic \bb tools, 
For the fully systematic process, the first author systematically searched for peer-reviewed papers in computer science academic databases. 
The less-systematic process was conducted using standard Google search queries.
In some cases, some tools reference other tools, so we also used this information.
Following this approach, we found 30 different relevant \bbes.

We consider using Google scholar for the systematic approach, but unfortunately, it provided more than 2.6k results, which is more than what we can deal with.
Therefore, we reduced the search space to the four primary academic databases in computer science and software engineering, namely, IEEE, ACM, Elsevier, and Springerlink.
 The selected academic databases are shown in \Cref{tab:table1}. 
 They were selected because these databases are well known, and they have proceedings of the leading journals and conferences on which \bbes\ have been applied, such as education, software engineering, human-computer interaction, and end-user programming.

\subsection{Queries}
To identify and understand languages that offer \bb editors, we used the following \textit{search string} in the academic databases:

%\centerline{\textit{\bb\ programming OR \bb\ languages.}}
\begin{quote}
\small
\texttt{block-based language OR blocks-based language OR block-based languages OR blocks-based languages OR block-based programming OR blocks-based programming}\\
%\vspace{-4mm}
\end{quote}
%https://scholar.google.com.co/scholar?start=210&q=%22block-based+language%22+OR+%22blocks-based+language%22+OR+%22block-based+languages%22+OR+%22blocks-based+languages%22+OR+%22block-based+programming%22+OR+%22blocks-based+programming%22&hl=en&as_sdt=1,5&as_vis=1
We used the search string mentioned above for all four academic databases.
A summary of the number of results obtained from each database is presented in \Cref{tab:table1}.
\Cref{tab:type} presents a summary of the type and number of publications obtained across all the databases.
The publication type \textit{Other} aggregates different types of publications such as demonstrations, posters, magazine columns, tutorials, outlines, living reference work entries, panels, conference description, editorials, and non-peer-reviewed technical reports. Details about the inclusion or exclusion criteria for the relevant proceedings are explained below in \Cref{ssec:incl}.
%\begin{table}[htp]
%\centering
%\begin{tabular}{l c}
%\toprule
%{\textbf{Source}} & {\textbf{\# Results}}\\
%  \midrule
%IEEE Xplore~\textregistered Digital Library  & 128\\
%The ACM Digital Library  & 272\\
%Elsevier ScienceDirect~\textregistered & 55\\
%Springerlink~\textregistered & 213\\ \midrule
%Total & 668\\\bottomrule
%\end{tabular}
%\caption{Number of publications obtained per academic database.}~\label{tab:table1}
%\end{table}
%
%\begin{table}[htp]
%\centering
%\begin{tabular}{l c}
%\toprule
%{\textbf{Publication type}} & {\textbf{\# Papers}}\\
%  \midrule
%Conference paper & 369\\
%Journal & 143 \\
%Other &	50\\
%Chapter & 44 \\
%Abstract & 40\\
%Short paper & 22\\\bottomrule
%\end{tabular}
%\caption{Number of publications per type.}~\label{tab:type}
%\end{table}

\begin{table}[htp]
\parbox{.45\linewidth}{
\centering
\footnotesize
\begin{tabular}{l c}
\toprule
{\textbf{Source}} & {\textbf{\# Results}}\\
  \midrule
IEEE Xplore~\textregistered Digital Library  & 128\\
The ACM Digital Library  & 272\\
Elsevier ScienceDirect~\textregistered & 55\\
Springerlink~\textregistered & 213\\ \midrule
Total & 668\\\bottomrule
\end{tabular}
\caption{Number of publications obtained per academic database.}~\label{tab:table1}
}
\hfill
\parbox{.45\linewidth}{
\centering
\footnotesize
\begin{tabular}{l c}
\toprule
{\textbf{Publication type}} & {\textbf{\# Papers}}\\
  \midrule
Conference paper & 369\\
Journal & 143 \\
Other &	50\\
Chapter & 44 \\
Abstract & 40\\
Short paper & 22\\\bottomrule
\end{tabular}
\caption{Number of publications per type.}~\label{tab:type}
}
\end{table}

\subsection{Inclusion and Exclusion Criteria}~\label{ssec:incl}
This section presents the criteria we used for both the systematic and the less-systematic approach.
\paragraph{Non-academic}
We included solely tools that can be used at the moment of the systematic review,
\begin{enumerate*}[label=(\roman*)]
	\item Open-source tools.
	\item Commercial tools with free trial.
\end{enumerate*}
This includes languages and tools that can be accessed only by contacting the authors, as described on the tool's website.
%Moreover, tools that can be executed in the testing machine. 
%The testing machine used for trying out the different tools and languages has the following characteristics: MacBook Pro (13-inch, 2017), processor 2,3 GHz Intel Core i5, Memory 8 GB 2133 MHz LPDDR3, running MacOS Mojave Version 10.14.6 (18G87).
\paragraph{Academic}

We  reviewed the title and abstract of each paper manually to remove all papers that certainly were not featuring  languages with \bbes. The proceedings used in this literature review are all peer-reviewed articles related to \bb programming in the broad sense, published between January 1st, 2005 and August 1st, 2020.
Note that we are interested in all articles related to \bb interfaces, so we included all articles that used or mentioned \bb languages or \bb programming even if they present applications or studies of the \bb metaphor solely.

We excluded articles on the following topics:
\begin{enumerate*}[label=(\roman*)]
	\item Visual languages that do not feature a block-based editor
	\item Studies not written in English
	\item Frame-based editing unless they provide a connection to \bb editors
	\item Data-flow programming
	\item Form-filling programming
	\item Wizardry metaphor~\cite{EsperFosterGriswold2013}
	\item Duplicate articles that present the same tool without adding a fresh perspective.
\end{enumerate*}

Finally, we excluded reference work entries, living reference work entries, and educational papers unless they introduce a new tool, a language, or an extension to an existing tool or language that uses a \bb editor.

\begin{figure}[t]
  \includegraphics[width=0.7\textwidth]{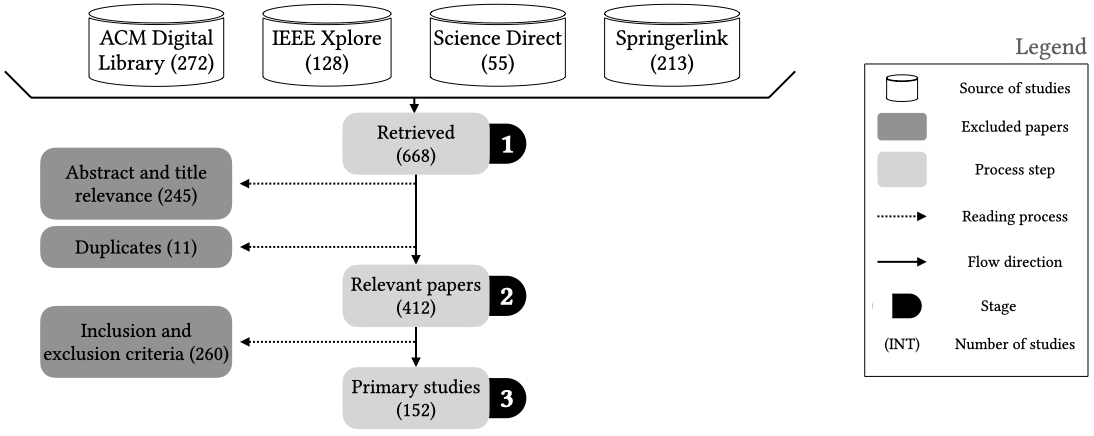}
  \caption{Selection procedure for the systematic literature review.}
  \label{fig:filtering}
%  \vspace{-4mm}
\end{figure}

\subsection{Selection}

To identify the relevant publications to be included as part of this SLR, we performed three-step filtering (\Cref{fig:filtering}) on the results obtained from all the databases using the string query mentioned before.
We took each result in the first filtering phase and we evaluated its relevance based on the title and the abstract only. Only papers that include something about \bbes where kept.
For each excluded paper, we wrote a motivation about why it was discarded.
After this process, the number of papers was reduced from 668 to 423.
After removing 11 accidental duplicates we ended up with 412 papers. The second filter step starts with all the papers that resulted from the first filter.
In this phase we defined nine yes/no questions based on our research questions. The nine questions are shown in \Cref{ap:questions}. Then we counted the number of yes answers for each individual paper.
Based on this count we chose a threshold to include a paper for the subsequent filtering step. 

Since answering the nine questions is a manual task, the second author double-checked a random selection of ten papers by following the exact same protocol. We measured the degree of agreement between both authors and we calculated \textit{Cohen's kappa} coefficient~\cite{cohen:1960}. This statistic is used to measure the degree of accuracy and reliability in statistical classification. Both authors agreed to include five papers and exclude four papers. However,  the first author decided to include one paper that the second did not. To quantify this: there was 90\% agreement between the authors and Cohen's kappa was 0.8. According to the guidelines proposed by Landis and Koch~\cite{Landis:1977}, a 0.8 Cohen's coefficient means that there is a \textit{substantial agreement} between the parties.

This literature review's primary focus is to provide a landscape of languages and tools related to \bbes. 
Therefore, the main criteria to include a paper is to introduce a language or a tool that uses a \bbe.
If that is the case, the paper is included even if the number of yes answers is not greater than the threshold.
If the paper does not include a language or a tool, we use the accumulated result to determine whether the paper is included.
Thus, a paper that does not introduce a tool or a language must have more than four positive answers.
As we did in the first filtering phase, we always record why a paper is discarded.

The second filter's resulting papers are then the ones on which the current survey is based; this means those are the papers from which we extract the data for further processing and discussion.
As a result of the second filtering phase, we excluded 260 papers from the 412 we had after the first filtering.
As a result we analyzed 152 papers in the data extraction phase.
During data extraction, we retrieved different elements, such as the type of publication, details about the \bbe\ (e.g., elements of the editor and its position on the screen), and all kinds of editor implementation details.
 All the data was collected in a spreadsheet and its content was then analyzed and processed by different means using scripts that aggregate the raw data. The result of this process is shown and explained in the following sections.

\section{Systematic review of \bbes}\label{sec:systematic}
In this section we answer the research questions (\Cref{sec:met}) using the data collected from the 152 papers on block-based editors.

\subsection{\qzero: \bzero}
This section presents demographics of the papers included in this survey.
Particularly, we present the venues in which the included papers were published, the number of papers included per year, and the number of papers per country.
\Cref{tab:venues} presents a summary of the venues that contributed the most number of papers.
For readability we present only the categories that contain venues that contributed at least two papers.
The complete list of categories and venues is listed in \Cref{ap:venues}.

\begin{table}[htp]
\centering
\footnotesize
\begin{tabular}{p{2.7cm} p{9cm} c}
\toprule
\textbf{Category}  & \textbf{Venue}       &  \textbf{\# Papers} \\
\midrule
\multirow{4}{*}{\begin{tabular}[c]{@{}l@{}}Human computer\\ interaction\end{tabular}} & Conference on Human Factors in Computing Systems (CHI) & 13 \\
 & Conference on Interaction Design and Children & 11 \\
 & International Conference on Human-Computer Interaction (HCII) & 4 \\
 & International Journal of Child-Computer Interaction & 3 \\
\midrule
\multirow{4}{*}{\begin{tabular}[c]{@{}l@{}}Programming /\\ Human computer \\ Interaction\end{tabular}} & Blocks and Beyond Workshop (Blocks and Beyond) & 13 \\
 & Symposium on Visual Languages and Human-Centric Computing (VL/HCC) & 9 \\
 & Journal of Visual Languages \& Computing & 2 \\
 & International Symposium on End User Development (IS-EUD) & 2 \\
\midrule
\multirow{7}{*}{Education} & Technical Symposium on Computer Science Education (SIGSE) & 10 \\
 & Global Engineering Education Conference (EDUCON) & 5 \\
 & Computational Thinking Education & 2 \\
 & Education and Information Technologies & 2 \\
 & Workshop in Primary and Secondary Computing Education & 2 \\
 & International Conference on International Computing Education Research & 2 \\
 & Conference on International Computing Education Research & 2 \\
\midrule
\multirow{1}{*}{Distributed computing} & International Conference on Computing, Communication and Networking Technologies & 5 \\
\midrule
\multirow{1}{*}{Robotics / Education} & International Conference on Robotics and Education (RiE) & 4 \\
 \midrule
\begin{tabular}[c]{@{}l@{}}Accessibility\end{tabular} & Conference on Computers and Accessibility (ASSETS) & 2 \\
\midrule
\multirow{1}{*}{Programming} & Science of Computer Programming & 2 \\
\midrule
Security & International Conference on Information Systems Security and Privacy (ICISSP) & 2 \\
\midrule
\multirow{1}{*}{Software engineering} & International Working Conference on Source Code Analysis and Manipulation (SCAM) & 2 \\
%\midrule
%Technology & International Conference on Advances in Information Technology & 1 \\
%\midrule
%\multirow{1}{*}{General} & Communications of the ACM & 1 \\
% \midrule
%\multirow{2}{*}{Engineering} & International Conference on Automation, Computational and Technology Management (ICACTM) & 1 \\
% \midrule
%\multirow{2}{*}{Robotics} & Iberian Robotics conference (Robot) & 1 \\
% & International Conference on Ubiquitous Robots (UR) & 1 \\
%\midrule
%eBusiness & International Conference on E-Business and Applications (ICEBA) & 1 \\
%\midrule
%Games & International Conference on Serious Games, Interaction, and Simulation (SGAMES) & 1 \\
%\midrule
%Electrical engineering & Computers \& Electrical Engineering & 1 \\
%\midrule
%Multimedia & Multimedia Tools and Applications & 1\\
\bottomrule
\end{tabular}
\caption{Summary of venues that contributed at least two papers to the survey.}
\label{tab:venues}
%\vspace{-9mm}
\end{table}

To get a quick overview of the most important venues we ordered them in \Cref{tab:venues} by ranking them by ``popularity''. Moreover, we manually classified them into 18 categories.
For the classification process we tried two semi-automated alternatives using a more systematic approach, namely (a) calculating the document distance between calls-for-papers of each venues and (b) using Google's Cloud Natural Language API~\footnote{https://cloud.google.com/natural-language} to classify each call-for-papers. The bottom-line is that both approaches did not produce accurate results and so we went back to the manual classification. We report on these negative results nevertheless, as they might be useful to others researchers that are working on an SLR.

We extracted the text in the call for papers of a random sample of venues to use these to test the two automated approaches. In this step, we notice that not all venues present a clear list of topics (e.g. the ACM CHI conference). For the first approach, we calculated the document distance between two calls for papers from the same field. By manually verifying documents which were either far apart, or close, with our own understanding we noticed nothing but noisy results. Apparently the variety of topics in calls for papers goes far beyond the variety of topics of what a conference is about.

To explore this further, we removed all the other text from the call from papers, and we calculated the document distances based only on the research topics mentioned in the call for papers. However, this did not improve the results, and the document distance between two venues from the same field was not too close (false negatives). And, in many cases even, comparing venues from distinct fields produced closer distances (false positives). 

The second approach used the same input data. We used the default Google's classification categories on the same texts, and the results were indeed accurate (correct), but they were not precise enough (vague). I.e. most of the venues were classified as ``computer science''.

After these failed attempts to automate and objectify our classification, we continued with a manual classification process.
\Cref{tab:venues} shows that the venues that contributed the highest amount of papers are CHI and `Blocks and beyond`, with 13 papers each.
The former is a venue about human factors in computing systems, including interaction, visualization, and human-computer interaction topics.
Thus, it is a clear connection between these topics and the benefits offered by \bbes.
The latter venue is exclusively focused on the development and use of \bbes.
Therefore it is a perfect match for the study we present in this survey.

The papers included in this study are from different domains such as Human-Computer Interaction (HCI), Education, Design, Software Engineering, Robotics, and Security.
Based on all the venues that contributed at least one paper, we expect our paper collection process to be rather complete for this study since we have publications from a variety of heterogeneous sources and topics.
Likewise, this study includes different types of proceedings as shown in \Cref{tab:type}.
% and the proceedings included in this study are of a variety of types, as shown in \Cref{tab:type}.

%Is my data of a good quality? Stats, etc.
%It’s complete, why it is good, etc. 
%Conference popularity.

%\begin{figure}[htp]
%  \includegraphics[width=0.5\textwidth]{assets/papers_per_year}
%  \caption{Summary of the number of papers published per year.}
%  \label{fig:papers_per_year}
%\end{figure}

\begin{table}[htp]
\parbox{.55\linewidth}{
\centering
\footnotesize
%\begin{figure}[htp]
  \includegraphics[width=0.55\textwidth]{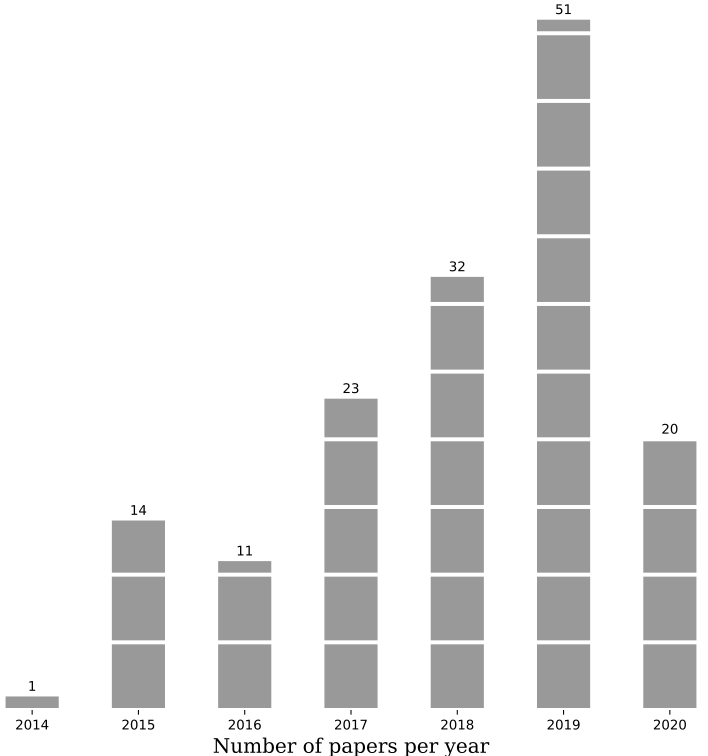}
  \caption{Summary of the number of papers published per year.}
  \label{fig:papers_per_year}
%\end{figure}

}
\hfill
\parbox{.35\linewidth}{
\centering
\small
\begin{tabular}{cc}
\toprule
\textbf{Publication Type} & \textbf{\# Papers} \\
\midrule
        Study &      31 \\
        Languages &      95 \\
        Extension &      27 \\
\bottomrule
\end{tabular}
\caption{Summary of the number of papers included in this study per type of proceeding}
\label{tb:type}
}
\end{table}

To understand the papers' demography, we computed the number of papers that we included in our study per year, as shown in \Cref{fig:papers_per_year}.
This figure shows that the number of papers per year has increased, having its peak in 2019.
It is important to remark that the current survey's search process solely included papers published before August 1st, 2020. 
This probably explains why the number of papers in 2020 is lower than in 2019.
With \Cref{fig:papers_per_year}, we can observe an increase in popularity on topics related to \bbes.

%\begin{figure}[htp]
%  \includegraphics[width=0.8\textwidth]{assets/papers_per_country}
%  \caption{This choropleth map displays a summary of the number of papers published per country. The detailed table with the values is shown in \Cref{tb:country}}
%  \label{fig:papers_per_country}
%\end{figure}

Moreover, we computed the number of papers published per country.
To compute this information we used the nationality of the first author as presented in the paper, and then we calculated the number of occurrences per country.
We can observe that the United States is the country that contributed the highest number of papers, followed by the United Kingdom with 64 and 14 papers, respectively.
It is essential to mention that the gap between the number of papers contributed by the US is more than four times the number of UK papers.
It is also interesting to observe that we have some degree of diversity in the authors' nationality; there are authors from different continents ---North America, South America, Europe, and Asia.  Antarctica, Africa, and Australia are not represented.
The complete list of papers per country is presented in \Cref{ap:countries}.

While analyzing the papers, we decided to tag them using three categories \textit{study}, \textit{language}, and \textit{extension}.
We defined these categories to classify the papers based on their content.
The first category, \textit{study}, is used to group papers that study aspects of using or implementing \bbes and do not present a new language or tool that uses the \bb metaphor.
The \textit{languages} category is used to group all the papers that present a new language that includes a \bb editor or tools that support the development of \bbes.
Finally, the \textit{extension} category groups papers that do not introduce a language but introduce new features to existing \bb editors.
\Cref{tb:type} presents a summary of the number of papers per category.

Based on the previous information, the reader can observe that the included papers come from a wide range of topics, types of publications, and authors from different parts of the world.
In the next section, we will present in more detail findings and information that we obtained by analyzing and gathering data from the corpus of papers, and that helps us answer the research questions defined in \Cref{ssec:research}.

\begin{tcolorbox}[colback=gray!5!white,colframe=black!75!black]
\footnotesize
\textbf{Summary \qzero}
\begin{itemize}
	\item Publication of \bbes is spread among different communities, however they are most present in education, human computer interaction, and programming venues.
	\item The number of publications that present \bb editors have been increasing since 2014. This is supported by the importance of programming in the last years among different people, including students and non-professional programmers.
	\item Authors from many countries publish papers that use \bbe. However, the country that contributes the most number of papers to this study is the United States, followed by the UK.
	\item In this survey, we classified the 152 papers based on their goal in three main categories, studies, languages, or extensions. Most of the papers included in this study are papers that introduce a language (95), followed by studies of the usage of \bb editors (31) and, finally, papers that introduce extensions to existing \bbes (27).
\end{itemize}
\end{tcolorbox}

\subsection{\qone: \bpone}\label{sec:qone}
% Feature diagram of a bbe
% Show stats about the components
% how many tools used them and where are they placed
This section addresses research question \qone based on the data collected.
For this purpose, we used the papers' classification from the previous section and we took the ones from the \textit{languages} group. From the total number of papers we considered a subset of 95 papers (\Cref{tb:type}).

\begin{figure}[t]
  \includegraphics[width=\textwidth]{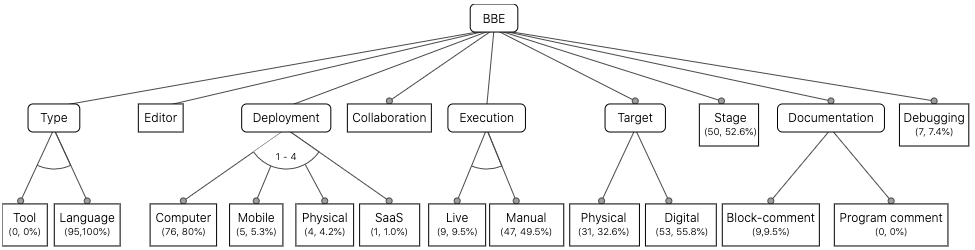}  
  \caption{This is the top-level of the feature diagram.}
  \label{fig:bbe}
\end{figure}

\begin{figure}[t]
  \includegraphics[width=\textwidth]{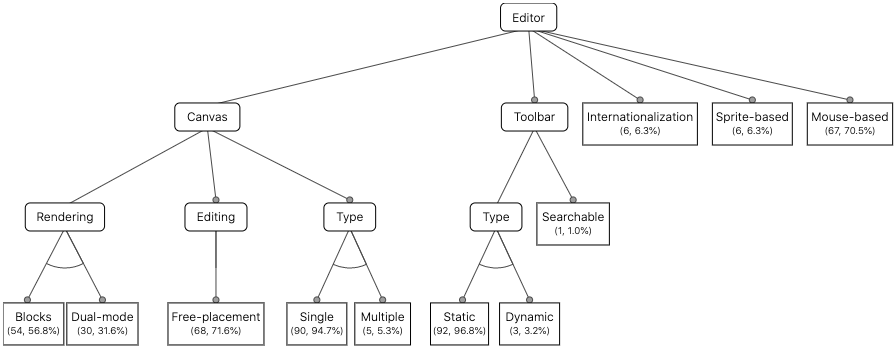}
  \caption{This feature diagram shows a zoom-in into the editor aspect of a BBE.}
  \label{fig:bbe_editor}
\end{figure}

Based on the different features offered from all the \bbes in this study, we developed a feature diagram~\cite{Kang:1990} that summarizes the most common features found across different platforms.
The complete set of features of \bbes is shown in \Cref{fig:bbe,fig:bbe_editor}.
To ease the diagram's readability, we split the \textit{editor} feature into a separate diagram, as shown in \Cref{fig:bbe_editor}.
\Cref{fig:bbe} shows the first part of the diagram.
Here the reader can observe features related to the functioning of the platform. 
For instance, \emph{licensing}, \emph{code execution mode}, and the \emph{type} of \bb environment.
Then, \Cref{fig:bbe_editor} presents details of the \bb editor.

In the feature model, we used two types of features, mandatory and optional.
The first is used for standard features (depicted as a box in \Cref{fig:bbe,fig:bbe_editor}), and the latter for unique features (depicted as a box with a blank circle on top).
The root node in \Cref{fig:bbe} represents a \bbe, and each of the leaf nodes in the feature diagram displays the number of \bbes that support that feature and the percentage of tools that support it among all the papers.
For instance, \emph{Computer (76, 80\%)} means that 76 \bbes are deployed for computers, which is 80\% of the papers used for creating this diagram.
All the \bbe's children nodes are described below.

\begin{description}
	\item [Type.] There are mainly two types of \bbes, \textit{tools}, and \textit{languages}. The former refers to utilities that help the development of such environments. Instead, the latter are languages that come with a \bb editor.
	\item [Editor.] \Bbes provide a \bb editor, but, we identified that some tools also support a hybrid editor (text and blocks), which means that it is possible to interact with the underlying language either through a blocks editor or text-based editor.
Based on this, 69 of the studied tools support a block editor only, while 15 support both blocks and text editor~\cite{Sutherland:2018,Bart:2017,devine:2019,Daehoon:2018,Beschi:2019,Benotti:2017,Kelly:2018,Blanchard:2019,Vostinar:2020,Leber:2019,Bachiller:2020,Alves:2020,Hussein:2019,Hernandez:2020,Ball:2019}.
The remaining ten tools do not mention it at all.
	
	\item [Deployment.]  A \bbe\ can be used through a heterogeneous set of devices (e.g., laptops, tablets, and wearables).
 Therefore, we investigated what device do \bbe\ users write or develop their programs with.
 The majority of tools are used through a browser-enabled PC (76), five through mobile devices (e.g., smartphones), four by manipulating physical elements, and one as Software as a Service (SaaS).
In nine of the tools, it was not clear which type of device the users have to develop their programs.
%	\item [Licensing.] We identified mainly three types of licensing in the papers and tools we studied, namely \textit{academic}, \textit{commercial}, and \textit{open souce}.
	
	\item [Collaborative.] This feature represents whether a \bbe\ supports mechanisms for users to collaborate in the development of programs.
From the studied papers, 90 tools do not offer such capabilities. 
Instead, the remaining five tools do support this feature.
	
	\item [Execution.] There are different ways of executing programs. This is not different in a \bbe.
Based on this study, we identified mainly two execution modes, \textit{manual} and \textit{live}.
After finishing the development of their programs, a manual execution means that users have to press a button to launch the execution of the program by the underlying language processors.
Instead, a \textit{live} execution mode does not require a manual intervention by the user to execute programs. 
The platform is capable of live executing the programs as users develop them.
From the tools, 47 use a manual execution mechanism, nine use a live execution, and the remaining 39 tools do not mention which execution mode do they use.
%	\item [Paradigms.] There are different programming paradigms~\cite{bb-paradigms} that can be supported or implemented by means of a language that offers a \bb editor namely \textit{configuration}, \textit{serial}, \textit{parallel}, and \textit{event-driven}.
	\item [Target.] As mentioned before, \bbes are used in different settings. This includes the real and the digital world.
Thus, sometimes the effects of running a \bb program are displayed on the screen, but sometimes they are shown via hardware, and sometimes both.
We investigated this fact and found that 53 tools present some form of results in a digital way, 31 using hardware (physical), and three using both.
For the remaining eight tools, it is not clear from the papers which one do they use.
	
	\item[Stage.] \Bb editors are used in different environments. In some environments, the effects of running a program are represented in the real world (e.g., hardware effects).
However, other contexts in which the results are displayed as software (e.g., animation). 
This feature characterizes if the \bbe\ has a dedicated pane for showing the effects of running a program.
Of the 95 tools, 50 of them have a dedicated pane for rendering program results; the remaining 45 do not have such a pane.
Since there is no standard way or location for placing a stage, we investigated the most common place in which \bbe\ developers place this component.
The preferred location for the canvas is the right-most side of the screen, with 22 tools. Then, 19 tools place it to the left-most part of the screen.
Two tools place it in the center; similarly, two tools have it in the bottom part.
Finally, only one tool has the stage on the top part of the screen.

However, the following four tools do have a stage, but the paper is not clear wherein the screen it is located.
Below, we present the details for these four tools.
Catrobat~\cite{Muller:2018} uses a different layout because it is a mobile app.
Behavioral Blockly~\cite{ASHROV2015268} does not show the whole \bbe. Some images show the programs, and others that show the stage, but not the whole workspace.
In VEDILS~\cite{Mota2018} there are different screens for showing the stage and editing the program.
For the mBlock~\cite{Lee:2020} tool it is not clear where the stage is located from the paper's screenshots.
	
	\item[Documentation.]
Documentation is an essential aspect of software.
In traditional text-based programming environments, it is possible to add comments anywhere in the program as long as it does not introduce syntactic errors.
In a \bbe, this is more restricted due to the projectional nature and the visual components. 
Therefore, we identified mainly two types of documentation. One is used to add documentation to specific blocks (\textit{block-comments}), and the other documents a complete program/script (\textit{program comments}).
	
	\item[Block-comments]
As introduced before, block-comments are the comments added to specific blocks. This could be either a group of blocks stacked up together or a single block.
From the studied papers, nine tools allow users to add comments per block, while the other 86 tools did not mention it explicitly.

	\item[Program comments.]
This feature is presented to show whether the tools allow users to add comments to complete scripts/programs.
We found that none of the tools found in this study support adding comments to \bb programs.
	
	\item[Debugging.] Traditional software development tools support the debugging of programs.
This is no different for \bbes; however, we found that not all \bbes support debugging features.
From all the tools, only seven tools come with debugging features.
The remaining 87 do not mention it; we assume they do not offer such capabilities.
\end{description}

Next, we present in detail the features that are part of the \bb editor \Cref{fig:bbe_editor}.

\begin{description}
	\item [Canvas.] The canvas is where users create their programs; it is where they drop the blocks that constitute programs.
All \bbes\ offer a canvas for building the programs.
Most of the papers (70) show their canvases, but some (24) papers did not present screenshots that show the canvas explicitly.
Some of these publications that did not present the canvas present \bb programs.
The canvas location indicates wherein the screen is this component situated.
For 49 of the tools, the canvas is located in the center of the window; 16 have it on the right-most part; one in the left-most part of the screen, and one have it in the bottom part.
As explained before, the remaining 27 tools do not mention or display their position.
	
	\item [Canvas type.] %As explained earlier, a canvas is a visual space in which users create their \bb programs.
Some environments provide more than a single canvas for creating programs. Therefore, we look at the papers, and we found that five tools do use multiple canvases, and the remaining 90 either only offer a single canvas or do not explicitly mention/show support for multiple canvases.
	
	\item [Rendering.] This feature means that the \bbe\ displays programs using only blocks or dual-mode (text and blocks).
54 of the \bbes display programs using only a block-based representation, 30 tools support a dual-mode, and the remaining 11 tools do not mention anything about it.
	
	\item [Editing.] A canvas allows users to build programs by placing blocks on it.
However, this does not mean that all \bb editors use a 2D space. 
From all the tools, the majority supports the free placement of blocks in a two-dimensional space.
However, the other 27 tools have other types of placement (e.g., 3D spaces or non-free placement of blocks).
	
	\item [Toolbar.] The toolbar is where blocks are grouped so that users can look at what language constructs (blocks) are available for further use.
Sixty-seven languages have a \bb editor that contains a palette, and 27 do not provide it or it is not explicitly mentioned.
Moreover, we analyzed the location of the palette also from the papers.
There are four possible locations \textit{top}, \textit{bottom}, \textit{left}, or \textit{right}.
We found that 47 tools have the palette on the left-most part of the window. This might be related that the majority of the people read from left to right.
Moreover, four tools (\textit{Flip}~\cite{Good:2017}, \textit{Labenah}~\cite{Alkadhi:2019}, ~\cite{lee:2019}, and \textit{Tuk tuk}~\cite{Koracharkornradt:2017}) have the palette in the right-most part of the screen.
Twelve tools have it in the middle of the screen; this behavior usually presents a stage on one side and the editor on the opposite side. In this way, the palette is in the middle. 
Finally, \textit{XLBlocks}~\cite{Jansen:2019} displays the toolbox at the top of the window and \textit{Tica}~\cite{Almjally:2020} does it in the bottom part.
	
	\item [Toolbar Type.] A palette usually groups blocks by categories and this grouping is \textit{static}, meaning users can inspect each category and its blocks, and it will not change.
However, we identified that some tools offer a \textit{dynamic} toolbar.
A dynamic toolbar is a toolbar that automatically adapts its contents based on the program's current status.
In other words, it automatically hides the blocks that cannot be snap into the current status of the program.
There are 91 tools that do not support this feature, but \textit{EduBot}~\cite{Islam:2019}, \textit{PRIME}~\cite{Rodriguez:2019}, and \textit{EUD-MARS}~\cite{Pierre:2020} do.
	
	\item [Searchable Toolbar.] A \textit{searchable palette} is a palette that has a search bar to help users find blocks without having to open each category. \textit{EduBot}~\cite{Islam:2019} is the only tool that supports a searchable toolbar.
	
	\item [Internationalization.] Given the visual notion of a \bbe\ and the possibility of adding descriptions to language constructs in natural language, we investigated if the \bb tools come with support for different languages, which means, if the description of a block can be shown in several languages (e.g., English, Spanish, Dutch).
We found that only six tools come with internationalization capabilities, and the vast majority (89) do not support it.
%	\item[Debugging.] Traditional software development tools support the debugging of programs.
%This is no different for \bbe, however, we found that not all \bbe support debugging features.
%From all the tools, only seven tools comes with debugging features.
%The remaining 87 either do not mention it, thus we assume they do not offer such capabilities.
	
	\item[Sprite-based.] Sprites are graphic elements of a computer program that can be manipulated as single units. This concept is popular among \bbe\ because Scratch supports it. 
However, we found that is not true for all languages that offer a \bb editor.
 We identified six tools that support first-class sprites, while the remaining 89 do not.

	\item[Mouse-based manipulation.] This feature is to reflect how users can manipulate blocks within a \bbe. 
Sixty-seven tools support the direct manipulation of blocks using the mouse, while the other 28 tools have different manipulation mechanisms (e.g., physical manipulation).
\end{description}

\begin{tcolorbox}[colback=gray!5!white,colframe=black!75!black]
\footnotesize
\textbf{Summary \qone}
\begin{itemize}
	\item The feature diagram (\Cref{fig:bbe,fig:bbe_editor}) displays the most important features across \bbes. There are features at two different levels, \textit{platform}, and \textit{editor}. At the \textit{platform} level, we find features such as documentation, collaborative support, deployment, and stage. The \textit{editor}-level features are the canvas, toolbar, internationalization, and sprite-based editing. Based on our data, we present quantitative analysis to illustrate which tools support each feature. Likewise, we also illustrate the position in which some of these features appear in a \bbe\ (e.g., canvas, toolbar, and stage location).
%	\item Based on the information collected from the papers, we identified that a standardized set of \bb editor features is missing. This is due to the different purposes for which these environments are used.
	\item We identified that --due to the diverse applications in which these environments are used-- a standardized set of \bb editor features is missing.	Therefore, we propose a feature diagram that summarizes them across different platforms. Notably, we identified two main types of features: platform-based and editor-based.
	\item We identified that most \bbes provide a palette that contains all the language construct and a canvas, in which users develop their programs. The stage is a key component in popular platforms, however, their presence varies depending on the language's goal.
	\item There are two main types of \bb editors: sprite-based (e.g., Scratch) and non-sprite-based.
\end{itemize}
\end{tcolorbox}

%\section{What are the tools used to develop \bbe s?}
\subsection{\qtwo: \bptwo} \label{sec:qtwo}
% How are BBE being developed
% based on the data show how many are implemented by hand
% enumerate the tools used and its frecuency e.g., scratch is used by x number of projects

We want to learn how \bbes are developed.
However, given the nature of the papers, this is a non-trivial activity because in most cases we noticed that authors do not mention these details. Below, we present the data we extracted.
Depending on how the language was implemented, we classified each paper into one of four categories General-Purpose Programming Language (GPL), grammar, DSL, and not available (N/A).
As shown in \Cref{tab:implementation}, 93 tools did not explicitly mention the tools used for its development, 55 were implemented using a GPL, and  from the remaining three: one used a visual language, one used a grammar, and one used a DSL, respectively.

%\begin{table}[htp]
%\centering
%\small
%\begin{tabular}{lc}
%\toprule
%\textbf{Category} &  \textbf{\# Languages} \\
%\midrule
%N/A                                   &                    93 \\
%GPL                                   &                    55 \\
%DSL                                   &                     1 \\
%Visual (blocks)                       &                     1 \\
%Grammar                               &                     1 \\
%\bottomrule
%\end{tabular}
%\label{tab:implementation}
%\caption{Language selection for the development of \bbes.}
%\end{table}

Likewise, we studied what programming languages were used in the implementation of these \bbes.
\Cref{tab:pls} presents a summary of our findings. For conciseness we grouped some of the languages (for the full list see \Cref{ap:pls}).
For instance, some languages only mention the use of HTML, so we count it as part of \textit{HTML, JavaScript, and CSS}.

%\begin{table}[htp]
%\centering
%\small
%\begin{tabular}{lc}
%\toprule
%\textbf{Programming language} &  \textbf{Papers} \\
%\midrule
%N/A                                        &                   100 \\
%JavaScript                                 &                    15 \\
%HTML, JavaScript, and CSS                  &                    15 \\
%Java                                       &                     3 \\
%Python                                     &                     2 \\
%iOS (Swift)                                &                     2 \\
%JavaScript \& Java                         &                     2 \\
%Pharo Smalltalk                            &                     2 \\
%TypeScript                                 &                     2 \\
%AngularJs and TypeScript                   &                     1 \\
%Snap!                                      &                     1 \\
%Python, HTML, CSS and Javascript           &                     1 \\
%Java / JastAdd                             &                     1 \\
%Java / ANTLR                               &                     1 \\
%ActionScript                               &                     1 \\
%PHP-based framework \& JavaScript and HTML5 &                    1 \\
%Tiled Grace                                &                     1 \\
%\bottomrule
%\end{tabular}
%\label{tab:pls}
%\caption{Programming languages used to implement \bbes.}
%\end{table}

\begin{table}[htp]
\parbox{.35\linewidth}{
\centering
\small
\begin{tabular}{lc}
\toprule
\textbf{Category} &  \textbf{\# Languages} \\
\midrule
N/A                                   &                    93 \\
GPL                                   &                    55 \\
DSL                                   &                     1 \\
Visual (blocks)                       &                     1 \\
Grammar                               &                     1 \\
\bottomrule
\end{tabular}
\caption{Language selection for the development of \bbes.}
\label{tab:implementation}
}
\hfill
\parbox{.55\linewidth}{
\centering
\small
\begin{tabular}{lc}
\toprule
\textbf{Programming language} &  \textbf{Papers} \\
\midrule
N/A                                        &                   100 \\
JavaScript                                 &                    15 \\
HTML, JavaScript, and CSS                  &                    15 \\
Java                                       &                     3 \\
Python                                     &                     2 \\
iOS (Swift)                                &                     2 \\
JavaScript \& Java                         &                     2 \\
Pharo Smalltalk                            &                     2 \\
TypeScript                                 &                     2 \\
%AngularJs and TypeScript                   &                     1 \\
%Snap!                                      &                     1 \\
%Python, HTML, CSS and Javascript           &                     1 \\
%Java / JastAdd                             &                     1 \\
%Java / ANTLR                               &                     1 \\
%ActionScript                               &                     1 \\
%PHP-based framework \& JavaScript and HTML5 &                    1 \\
%Tiled Grace                                &                     1 \\
\bottomrule
\end{tabular}
\caption{Programming languages used to implement \bbes.}
\label{tab:pls}
}
%\vspace{-4mm}
\end{table}
% TODO: Thesis only

 As mentioned before, implementation details are not always discussed, and this is reflected in \Cref{tab:pls}; 100 papers do not mention what programming language was used for the development.
 After this, we see that the most popular programming language for the development of \bbes\ is JavaScript.
Counting all the appearances, this language was used in the development of more than 30 block-based editors.
Another interesting fact is that there is only one language developed using a Language Workbench (JastAdd~\cite{Techapalokul:2019}).

Following this direction, we explored whether the papers did not mention programming languages at all, or it was just that they did not present implementation details of their tooling.
We used the list of the 50 most popular languages as reported by the TIOBE index~\cite{tiobe:2021}, but ``visual basic'' was omitted from the search because of the many false positives with the common words ``visual'' and ``basic''. In fact we did not find any block-based editor that was implemented in Visual Basic.

\begin{figure}[htp]
  \includegraphics[width=\textwidth]{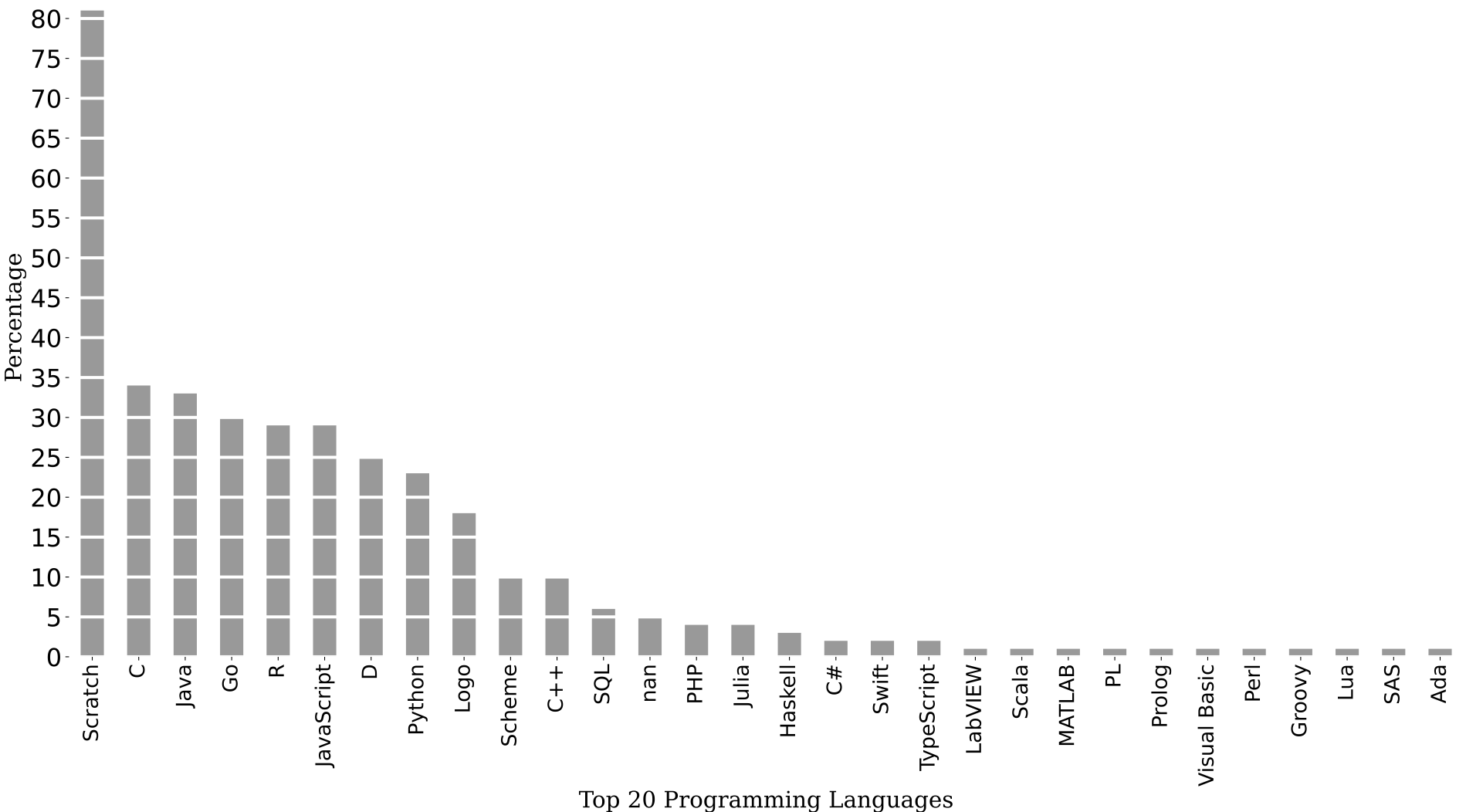}
  \caption{Summary of programming languages used for implementing \bbes.}
  \label{fig:pls_perc}
\end{figure}

Based on the list of programming languages, we developed a tool~\footnote{\url{https://github.com/maveme/PDFMiner}} for mining the corpus of PDF files and counting the occurrences of each programming language.
The results in \Cref{fig:pls_perc} show the popularity of each of programming language.
The complete list of details of each language and the number of papers that mention the language is presented in \Cref{ap:pls}.

As shown in \Cref{fig:pls_perc}, Scratch is by far the language most mentioned across the papers.
The reason for this is that most of the current \bbes took inspiration from it.
Then, we found seven programming languages (C, Java, Go, R, JavaScript, D, and Python) mentioned in more than 20\% of the papers.
These languages' popularity might be related to the technologies used to develop \bbes, and the libraries offered to support their development (e.g., Blockly).

%However, there is no clear reason about the popularity of all these seven languages in the paper, but this results show that they follow the trend presented in the TIOBE index.

% this is a conlcusion
In summary, we identified that most of the papers do not present implementation details about their languages and editors.
However, based on the papers that present implementation details, we found that most of the authors use GPLs.
Concretely, most of the papers that presented such details used HTML, JavaScript, and CSS to implement \bbes.
Likewise, we observed that the programming languages used to develop \bb editors are aligned with the 50 most popular languages as classified in the TIOBE index.

%
%This list is based on the 20 most popular languages from the TIOBE index\cite{tiobe:2021}. 

\begin{tcolorbox}[colback=gray!5!white,colframe=black!75!black]
\footnotesize
\textbf{Summary \qtwo}
\begin{itemize}
	\item We identified different ways in which \bbe\ are developed. However, most of the authors (93) did not include such details. The most popular way of developing a \bbe\ is employing a general-purpose programming language (GPL).
	\item Since using a GPL is the most common way of developing \bbe, we identified that the most popular languages for this endeavor are HTML, JavaScript, and CSS.
	\item Language Workbenches are really under-represented as a means of implementing a block-based editor. There seems to be an opportunity there. 
\end{itemize}
\end{tcolorbox}

%\section{How are \bbe s developed?}
\subsection{\qthree: \bpthree}\label{sec:qthree}
One of our main objectives of this mapping study is to identify how \bbes are developed in practice.
Therefore, we searched the selected papers for the languages and tools used by the authors to develop \bbes.
Based on the data collected (see \Cref{tab:how}) we identified two ways of implementing a \bb editor: either authors rely on existing \textit{libraries and frameworks} or they develop them in a \textit{bespoke} fashion.
From the corpus of papers that presented a language, tool, or an extension, 88 of them used libraries for the development of their editors, nine papers developed their bespoke editors entirely from scratch, and 54 papers did not provide a clear insight about how they were implemented, or they did not necessarily introduce a new tool.
However, to better understand of how \bbes are developed, we analyzed the papers to extract the libraries and frameworks used for their development.

%The whole set of tools/libraries used is shown in \Cref{tab:meta-tools}.

%\begin{table}[htp]
%\centering
%\small
%\begin{tabular}{lc}
%\toprule
%\textbf{Method} &  \textbf{\# Languages} \\
%\midrule
%Libraries and Frameworks              &             88 \\
%N/A                                   &             54 \\
%Bespoke                                &              9 \\
%\bottomrule
%\end{tabular}
%\label{tab:how}
%\caption{Method used for developing \bbes.}
%\end{table}

\Cref{tab:meta-tools} shows a summary of the resulting list of tools and libraries (for the full list see \Cref{ap:devtools}).
Similarly, as identified in \Cref{sec:qtwo}, most of the papers do not disclose implementation details.
However, due to the popularity of some tools for building \bb editors (such as Blockly) in some cases we were able to identify which library was used for their development, even if the authors did not mention them.

\begin{table}
\parbox{.35\linewidth}{
\centering
\small
\begin{tabular}{lc}
\toprule
\textbf{Method} &  \textbf{\# Languages} \\
\midrule
Libraries and Frameworks              &             88 \\
N/A                                   &             54 \\
Bespoke                                &              9 \\
\bottomrule
\end{tabular}
\caption{Method used for developing \bbes.}
\label{tab:how}
}
\hfill
\parbox{.55\linewidth}{
\centering
\small
\begin{tabular}{lc}
\toprule
\textbf{Library} & \textbf{\# of languages}  \\
\midrule
N/A                                &              62 \\
Blockly                            &              40 \\
Scratch                            &              11 \\
Snap!                              &               7 \\
Scratch 3.0 (Blockly)              &               3 \\
CT-Blocks                          &               3 \\
App inventor \& Blocky             &               2 \\
Microsoft MakeCode                 &               2 \\
BlocklyDuino                       &               2 \\
%Python                             &               1 \\
%NetTango                           &               1 \\
%Blockly@rduino                     &               1 \\
%Blockly and Standard pictogramming &               1 \\
%Nettango                           &               1 \\
%Blockly, Monaco, MakeCode          &               1 \\
%iSnap~\cite{Price:2017}            &               1 \\
%Droplet                            &               1 \\
%MakeCode                           &               1 \\
%Snap! \& Scratch                   &               1 \\
%Deltatick                          &               1 \\
%NetsBlox                           &               1 \\
%PencilCode, Droplet                &               1 \\
%Openblocks~\cite{Roque:2007}       &               1 \\
%Ardublock                          &               1 \\
%Tiled Grace                        &               1 \\
%Scratch \& Blockly                 &               1 \\
%Snap! \& DB Snap                   &               1 \\
%App inventor                       &               1 \\
\bottomrule
\end{tabular}
\caption{List of tools used for the development of \bb editors.}
\label{tab:meta-tools}
}
%\vspace{-8mm}
\end{table}
% TODO: only for thesis

As we can observe in \Cref{ap:meta-tools}, there are more than 20 libraries or frameworks used by authors.
The most popular tool used for developing \bbes is Blockly. It is one of the few tools specifically designed to support the development of \bb editors, which explains its observed popularity.
Moreover, it is interesting to observe that some of the tools used for building the languages are also \bbes (e.g., Scratch, Snap!), which means developers rely on existing languages and editors for the development of \bbes.
This is interesting and worth studying in the future, perhaps there is a lack of specialized tools for building \bbes, or simply the Software Language Engineering (SLE) community is not aware of the opportunities offered by \bbes.

%\begin{figure}[htp]
%  \includegraphics[width=\textwidth]{assets/pls_perc_50}
%  \caption{Popularity summary of \bbe across publications.}
%  \label{fig:pls_perc}
%\end{figure}

\begin{tcolorbox}[colback=gray!5!white,colframe=black!75!black]
\footnotesize
\textbf{Summary \qthree}
\begin{itemize}
	\item Most of the \bb editors included in this survey were developed using libraries and frameworks; only nine editors were developed in a bespoke fashion.
	\item More concretely, the most popular libraries used for developing \bbes are Blockly and Scratch.
\end{itemize}
\end{tcolorbox}

%\section{What \bb languages exist and what are they being used for? Where are \bbe being used?}
\subsection{\qfour: \bpfour}\label{ssec:qfour}
As part of this systematic literature review, we sought for the usages of \bbes.
This means, understanding the existing languages that support a \bb editor, and how these languages are being used.
While talking to colleagues we noticed that there is a perception that the \bb notation is restricted to computer science education.
People also seem to associate \bbes with children's tools or toys, given their colorful appearance.
To check the validity of these perceptions we analyzed the corpus of papers and  documented what tools are introduced in each paper and in which fields these tools are used.
%For this purpose, we defined 12 categories, in which we classified each paper.

The process to extract this data from the papers is described below.
First, during the paper review, we collected specific notes in a spreadsheet about each tool. 
We noted down a possible topic for each tool. Then, with the resulting data, we calculated the number of topics.
Initially, we obtained 81 topics, but that classification was not accurate enough to group the papers in a meaningful manner. 
Thus, to reduce this number and make a more accessible grouping of papers, we defined seven categories: Education, Embedded Computing, Human Computer Interaction, Arts \& Creativity, Science, AI, Data Science and Databases,  and Software Engineering.
% TODO: Thesis only %(\Cref{tab:categories}).
This number is significantly lower, and it works appropriately for presenting our findings.

As mentioned previously, we classified the papers into three categories, namely \textit{Research}, \textit{Language}, or \textit{Extension}; and the way we present them in this section differs depending on their type.
%We made a distinction on how to present the information between research papers and the other two categories.
Research papers are presented with a summary that contains the paper's topic; and \textit{Languages} and \textit{Extensions} are summarized in a table containing the name of the language/extension, the library used for its development, and its primary usage.

%Below, we present each category in detail; and each of them might contain a combination of these three types.

%However, the way they a paper is presented differs depending on the type of publication.

%As a result, we grouped the 81 papers into 7 categories as shown in \Cref{ap:domains}.
%We use these categories (\Cref{tab:categories}) to present the languages and the extensions we found in the papers.
%Next, we present each of the categories divided by topics, and each topic presents the tool/language introduced.

%\begin{table}[htp]
%\centering
%\footnotesize
%\begin{tabular}{c}
%\toprule
%\textbf{Category} \\
%\midrule
%Education\\
%Embedded Computing \\
%Human Computer Interaction\\
%Arts \& Creativity\\
%Science \\
%AI, Data Science, and Databases\\
%Software Engineering\\
%\bottomrule
%\end{tabular}
%\caption{Grouping categories identified in the papers.}
%\label{tab:categories}
%\end{table}

% TODO: Thesis only

Below we present each category with a brief description and a table with the papers that belong to it.
However, to improve the readability of the current manuscript, some categories have more than one table.

\begin{mybox2}[Education]
\small
This category presents the papers that are mostly related to educational purposes.
There is a wide range of applications in which \bbes are used to teach programming or computer science concepts, and other subjects such as chemistry and mathematics.
Likewise, this section presents the importance of \bbes in educational settings and how the modality (blocks or text) affects the learning process.
\end{mybox2}

\subsection*{Programming literacy}
In this category, we grouped several topics and points of view regarding using of the \bb modality in programming education.
For instance, Weintrop~\cite{Weintrop:2019} studied the impact of using a \bbe\ in education, and Xinogalos et al.~\cite{Xinogalos2017} investigates students’ perceptions of five popular educational programming environments and the features that introductory programming environments must-have.
Similarly, Yoon et al.~\cite{Yoon2019} designed a curriculum that integrates socio-scientific issues in the design and development of mobile apps using App Inventor.
Turbak et al.~\cite{Turbak:2014} studied the importance of teaching event-based programming in computer science curricula.
Dong et al.~\cite{Dong:2019} propose a tinkering environment for students when they struggle in problem-solving activities.
Dwyer et al.~\cite{Dwyer:2015} study the readability of \bb programs by students.

There are different points of view regarding the modality in which programming should be introduced to novice users.
Some advocate that visual languages are the best option for introducing novices to programming, while others support text-based languages as the best modality since that is what professional developers use.
Thus, researchers have tried to address this topic, and they have work on evaluating the effects that the modality (block-based, text-based, and hybrid) has in the learning process~\cite{Weintrop:20183,Brown:2016,Price:2015,Weintrop:20172,Mladenovic2018}.
Franklin et al.~\cite{Franklin:2016} study the differences between \bb languages (e.g., Scratch) and text-based languages (e.g., C and Java).
Other researchers focus on studying how to ease the transition from a \bb language into a text-based language~\cite{Kolling:2015, Weintrop:20192} and the drawbacks users face in this transition ~\cite{Moors:2017, Moors:2018}.
% accesibility
Milne and Ladner~\cite{Milne:2019} study the relevance of accessibility features in \bbes.
% transitioning from block-text 

%\cite{Brown:2016} shows that the modality used to introduce programming concepts matter.

%\cite{Kolling:2015} proposes frame-based editing as a solution to ease the transition from \bbe s to text-based environments.
%\cite{Price:2015} compare text vs block notations when learning programming
%\cite{Weintrop:20192} transitioning from \bbe and text-based environments to professional programming languages in high school.
%\cite{Mladenovic2018} comparing misconceptions in loops constructs in \bbe and text-based languages.
%\cite{Moors:2017} described problems novices experience when transitioning from \bb to traditional programming environments.
%\cite{Weintrop:2017} investigates the affordances of hybrid block/text based programming environments in high school.
%Moors et al.~\cite{Moors:2018} describe and present the drawbacks observed in students when moving from a \bbe into a text-based environment.

%\subsection*{Programming literacy: Languages}

Finally, \Cref{tab:programming-edu} presents the tools aimed at teaching computer science concepts in general and learning environments to support the teaching of computational concepts.

\begin{table}[htp]
\footnotesize
\begin{tabular}{l >{\centering\arraybackslash} m{2cm} l}
\toprule
{\textbf{Name}} & {\textbf{Metatool}} & {\textbf{Topic}}\\\midrule
% teach programming/concepts
MUzECS~\cite{Bajzek:2015} & Blockly & Explore computer science with a low-cost robot.\\
RoboScape\cite{Ledeczi:2019} & NetsBlox & Teach key ideas in computer science using robots.\\
\cite{Vostinar:2020} & MakeCode & Foster computer science education with Lego Mindstorms.\\
Robot Blockly~\cite{Weintrop:2017} & Blockly & Programming industrial robot (ABB’s Roberta).\\
HIPE\cite{Daehoon:2018} & - & Pedagogy and programming education.\\
Reduct~\cite{Arawjo:2017} & - & Gamifying operational semantics.\\
\cite{Seraj:2019} & Blockly & Introduce young learners to technology with smart homes.\\
Labenah\cite{Alkadhi:2019} & - & Learn coding principles via an edutainment application. \\
Bubbles\cite{Reinhardt:2017} & Ardublock & Teach programming to children with a robot fish.\\
Blocks4DS\cite{Feijoo:2019} & Blockly & Teach data structures. \\
Crescendo\footnotemark~\cite{Wang:2020} & Snap! & Engage students with programming.\\
%Parallel 
\cite{Feng:2015,Feng:2016,Feng:2017} & Snap! & Add parallel abstractions to \bb languages. \\
% using games
Pirate Plunder~\cite{Rose:2019} & - & Teach abstractions and reduce code smells with a game.\\
Resource Rush\cite{Lytle:2019} & Blockly & Teach programming in an open-ended game environment.\\
Block-C\cite{Kyfonidis:2017} & Openblocks & Learn the C language. \\
Cake~\cite{Jung:2016} & Blockly &  Learn the C language.\\
% learning environment
Block Pictogramming~\cite{Ito:2020} & Blockly & Learn programming with pictograms.\\
PRIME~\cite{Rodriguez:2019} & Blockly & Learning environment.\\
Flip~\cite{Good:2017} & - & Learn computer science in a bimodal environment.\\
IntelliBlox~\cite{Taylor:2019} & Blockly & Introduce programming in game-based environments.\\
EduBot~\cite{Islam:2019} & Blockly & Learn programming and STEM modules. \\
Alg-design\textsuperscript{\ref{exte}}~\cite{Vinayakumar:2018} & CT-Blocks & Teach algorithmic to novices.\\
Map-Blocks~\cite{Vinayakumar:20184}  & CT-Blocks & Teach programming with online weather data.\\
% design \bbe for middle school
LAPLAYA~\cite{Hill:2015} & Snap! & \Bbe\ for middle school students.\\
\bottomrule
\end{tabular}
\caption{Languages used to support programming education.}
\label{tab:programming-edu}
%\vspace{-4mm}
\end{table}

\footnotetext{This \bbe\ is not a complete platform but an extension to an existing system.\label{exte}}

\Cref{tab:ct} shows the languages used to support and transfer computational skills to learners.

\begin{table}[htp]
\footnotesize
\begin{tabular}{l >{\centering\arraybackslash} m{2cm} m{8cm}}
\toprule
{\textbf{Name}} & {\textbf{Metatool}} & {\textbf{Topic}}\\\midrule
% Computational thinking
PiBook~\cite{Campos:2018} & Blockly & Transfer computational skills while working on history, biology, and mathematics. \\
TunePad~\cite{Gorson:2017} & Blockly & Introduce computational thinking via sound composition.\\
C3d.io~\cite{madar:2019} & Blockly & Create 3D environments to enable STEAM education. \\
Tuk tuk~\cite{Koracharkornradt:2017} & - &  Teach computational thinking concepts using games. \\
CT-Blocks~\cite{Vinayakumar:20182} & - & Teach computational thinking skills. \\
ChoiCo~\cite{Grizioti:2018} & Blockly & Teach computational thinking via modifying games.\\
\cite{lee:2019} & - & Teach computational thinking.\\
\cite{Vinayakumar:20183}\textsuperscript{\ref{exte}} & Scratch & Teach computational thinking using experiments of fractal geometry.\\
\bottomrule
\end{tabular}
\caption{Languages used to teach computational thinking skills.}
\label{tab:ct}
%\vspace{-4mm}
\end{table}

\Cref{tab:others} contains the \bb languages used to teach other subjects such as aerodynamics, Latin language, mathematics, music, and chemistry.

\begin{table}[htp]
\centering
\footnotesize
\begin{tabular}{m{3cm} c m{8cm}}
\toprule
{\textbf{Name}} & {\textbf{Metatool}} & {\textbf{Topic}}\\\midrule
% teach aerodynamic and programming
Airblock~\cite{Breuch:2020} & Scratch & Teaching programming and aerodynamics.\\
% networks
BlockyTalky~\cite{KELLY20188} & Blockly & Teaching networks.\\
% teach grammar of the Latin language
Ingenium~\cite{Zhou:2016} & Blockly & Teaching Latin grammar.\\

% application
%simulation based
ExManSim~\cite{Rouwendal:2020} & Blockly & Create vignettes for crisis response exercises.\\

% Mobile apps development
Catrobat~\cite{Muller:2018} & - & Develop mobile applications collaboratively.\\
MIT App Inventor~\cite{Patton2019} & Blockly & Develop mobile applications.\\

% Agent-based simulations
EvoBuild\textsuperscript{\ref{exte}}~\cite{Wagh2018} & Deltatick\cite{WILKERSON:2015} & Teach and create agent-based models. \\
Phenomenological gas particle sandbox\cite{Aslan:2020} & NetTango & Teach agent-based computations through phenomenological programming.\\
% chemistry
M.M.M.~\cite{Saba:2020} & Blockly & Create an agent-based modeling system to learn science.\\
\cite{Bain:2019} & NetTango & Use agent-based modeling for other disciplines (e.g., chemistry).\\
% math
ScratchMaths\textsuperscript{\ref{exte}}~\cite{BENTON201868} & Scratch & Understand mathematical concepts through programming activities.\\
\cite{Kong2019}\textsuperscript{\ref{exte}} & App inventor & Teach mathematical concepts in primary school. \\
Tactode\cite{Alves:2020} & - & Teach math and technology concepts to children.\\

%music
Sonification Blocks~\cite{Atherton:2017} & Blockly & Learn data sonification.\\
\bottomrule
\end{tabular}
\caption{Languages used to teach subjects other than programming.}
\label{tab:others}
%\vspace{-8mm}
\end{table}

\Cref{tab:more} presents tools aimed to improve the transition from \bb languages to text-based languages, incorporating \bb notation to existing environments such as spreadsheets, languages to support teachers during grading activities, and, finally, languages to support learners with accessibility issues (e.g., hearing impairments).

\begin{table}[htp]
\footnotesize
\begin{tabular}{m{3cm} >{\centering\arraybackslash} m{2.5cm} l}
\toprule
{\textbf{Name}} & {\textbf{Metatool}} & {\textbf{Topic}}\\\midrule
% Block-text
Amphibian~\cite{Blanchard:2019} & Droplet & Enable switching between blocks and text.\\
Poliglot~\cite{Leber:2019} & Blockly & Smooth transition from blocks to text in education.\\
HybridPencilCode~\cite{Hussein:2019} & PencilCode and Droplet & Transition from block to text notation.\\
B@ase~\cite{Tamilias:2017} & Blockly@rduino & Transition from block to text-based environment. \\
PyBlockly~\cite{Strong:2018} & Blockly & Add a \bb editor for Python.\\
% frame-based
Stride~\cite{Kolling:2019} & - &  Add a frame-based editing (blocks and text) to BlueJ.\\
XLBlocks~\cite{Jansen:2019} & Blockly & Add \bbe\ for spreadsheets.\\
% Teacher support
NoBug’s SnackBar\cite{Vahldick:2017}  & - & Measure students' performance in programming tasks.\\
GradeSnap\cite{Milliken:2020} & Snap! & Assist teachers in grading \bb projects.\\

% accessibility
StoryBlocks\cite{Koushik:2019} & - & Teach programming to blind users with a tangible game.\\
Blocks4All~\cite{Milne:2018}  & - & Accessibility support for \bbes.\\
\cite{ong:2019}\textsuperscript{\ref{exte}} & Blocks4All & Accessibility support for \bbes.\\
Macblockly\textsuperscript{\ref{exte}}~\cite{Caraco:2019} & Blockly & \Bb support for audiences with disabilities. \\
\cite{Meenakshi:2020} & Blockly & Support users with hearing impairments to learn programing.\\

\bottomrule
\end{tabular}
\caption{\Bb languages applications.}
\label{tab:more}
\end{table}

%\subsection{Embedded Computing}

\begin{mybox2}[Embedded computing]
\small
This category contains all the papers that were associated with some form of embedded computing.
This includes languages for programming and manipulation of robots, microcontrollers, and other embedded systems.
\end{mybox2}

Following the idea of embedded computing with a \bbe, ~\cite{Capay:2019} present the benefits of using a \bb language for manipulating and teaching physical components.

\Cref{tab:embedded} presents all the languages we classified as being part of the embedded computing category.
This includes programming robots, embedded systems, Internet of Things (IoT) devices, and controllers.

\begin{table}[htp]
\footnotesize
\begin{tabular}{m{3.2cm} >{\centering\arraybackslash} m{2.7cm} l}
\toprule
{\textbf{Name}} & {\textbf{Metatool}} & {\textbf{Topic}}\\\midrule

MakeCode~\cite{Ball:2019} & Blockly & Programming environment for microcontrollers.\\
NaoBlocks~\cite{Sutherland:2018} & Blockly & Manipulate Nao robots.\\
Coblox~\cite{Shepherd:2018} & Blockly & Programming ABB's industrial robots.\\
ROS educational\textsuperscript{\ref{exte}}~\cite{Tatarian:2019} & Snap! & Manipulate ROS-enabled platforms. \\
Robobo~\cite{Bellas:2018,Bellas:2020} & Scratch & Manipulate advanced sensors. \\
%\cite{Bellas:2018} & Robobo & Scratch and Blockly & Case study & Programming environment to manipulate advanced sensors \\
EUD-MARS~\cite{Pierre:2020} & Blockly & Use model-driven approach to program robots.\\
CoBlox~\cite{Weintrop:20184} & Blockly & Interface for Roberta a single-armed industrial robot. \\
MakerArcade~\cite{Seyed:2019} & MakeCode & Create gaming experiences through physical computing. \\

% I think the main purpose is education through robots
UNC++Duino~\cite{Benotti:2017} & BlocklyDuino & Program robots to teach CS concepts.\\
The Coffee Platform~\cite{Sarmento:2015} & Blockly & Support computational thinking skills through robotics.\\
LearnBlock~\cite{Bachiller:2020} & - & Robot-agnostic educational tool.\\
RP~\cite{Avraam:2020} & Blockly & Affordable robot (software and hardware) for education.\\
mBlock~\cite{Lee:2020} & - & Teach CS and electronics with affordable robots.\\

% Collaborative
CAPIRCI~\cite{Beschi:2019} & - & Support collaborative robot programming. \\

%embedded systems
CODAL~\cite{devine:2019} & Blockly and MakeCode & Create effective and efficient code for embedded systems.\\

%Testing
OPEL TDO~\cite{Koehler:2018} & Blockly & Test programmable logic controllers by end-users. \\

%smart cities/big data
Block-based data fusion~\cite{Bonino:2016} & - & Define complex event processing pipelines for smart cities.\\

\bottomrule
\end{tabular}
\caption{\Bb languages in embedded computing.}
\label{tab:embedded}
%\vspace{-8mm}
\end{table}

%\subsection{Human Computer Interaction (HCI)}
\begin{mybox2}[Human Computer Interaction (HCI)]
\small
This category contains papers that focus on a wide variety of aspects of Human-Computer Interaction.
We identified aspects such as the usability of \bbes and their limitations, comparison between different user interfaces (e.g., TUIs, GUIs, and brain-computer interfaces), adding code hints to \bbe, and supporting end-user development (EUD) through \bb languages.
% usability
% interfaces TUIS vs GUIS, Brain
% hints
% end-user development
% story telling
%
\end{mybox2}

Most of the papers that fall in this category present a language as summarized in \Cref{tab:hci}.
However, three papers present a more theoretical view.
For instance, Holwerda and Hermans~\cite{Holwerda:2018} present an evaluation to measure the usability of Ardublockly~\cite{ardublockly}, a \bbe\ for programming Arduino boards.
This evaluation was done using the cognitive dimensions of notations framework ~\cite{CDN:2001}.
Likewise, Rough and Quigley~\cite{Rough:2019} present the challenges of traditional usability evaluations.
Almjally et al.~\cite{Almjally:2020} present an empirical study that compares the usage of a \bb language using Tangible User Interfaces (TUIs) and Graphical User Interfaces (GUIs).

% usability
%Holwerda and Hermans~\cite{Holwerda:2018} presents a usability evaluation of \bbe (Ardublockly~\cite{ardublockly}) using the cognitive dimensions of notations framework \cite{CDN:2001}.
%Rough and Quigley~\cite{Rough:2019} presents the challenges of traditional usability evaluations

%Almjally et al.~\cite{Almjally:2020} empirical study that compares Tangible User Interfaces (TUIs) and Graphical User Interfaces (GUIs) with a \bb language

\begin{table}[htp]
\footnotesize
\begin{tabular}{l >{\centering\arraybackslash} m{2cm} m{8cm}}
\toprule
{\textbf{Name}} & {\textbf{Metatool}} & {\textbf{Topic}}\\\midrule
% Usability of bbe
Shelves\textsuperscript{\ref{exte}}~\cite{Hsu:2017} & Blockly & Usability of \bbe.\\
%\cite{Rough:2019} & Jeeves &  & Case study & \bbe for defining time and context-based triggers upon to execute actions \\

% improve bbe hints
\cite{Marwan:2019}\textsuperscript{\ref{exte}} & iSnap!\cite{Price:2017} & Improve code hints in \bbe.\\
iSnap\textsuperscript{\ref{exte}}~\cite{Price:20172} & Snap! & Add intelligent tutoring system features to Snap!.\\

% interaction
\cite{Almusaly:2018} & - & Add custom keyboard to \bb languages.\\
Enrect~\cite{Suzuki:2018} & - & Introduce noted-link interfaces to represent variables.\\
Multi-device Grace~\cite{Selwyn-Smith:2019} & Tiled Grace & Support for multi-device environments.\\
% tangible
ARcadia~\cite{Kelly:2018} & MakeCode & Prototype tangible user interfaces.\\

% Teaching/education
VEDILS~\cite{Mota2018, Mota:2018} & App inventor and Blockly & Support end-users to create mobile learning applications with augmented reality. \\
%\cite{} & VEDILS & App inventor and Blockly & Case study & Framework to create mobile application for users without programming skills \\
Jeeves~\cite{Rough:2020} & - & Support end-users to develop applications.\\
% end-user programming
TAPAS~\cite{TURCHI201766} & - & Create workflow applications (e.g., IFTTT~\cite{IFTTT}). \\
TAPASPlay\textsuperscript{\ref{exte}}\cite{Turchi2019} & TAPAS & Support EUD via collaborative game-based learning.\\

% Story telling / IoT
StoryMakAR~\cite{Terrell:2020} & BlocklyDuino & Support storytelling with augmented reality and IoT.\\
Aesop~\cite{Saini:2019} & - & Create digital storytelling experiences.\\

%brain computer interfaces
Neuroblock\cite{Crawford:2019,Crawford:2018} & Scratch & Build applications driven by neurophysiological data.\\
%\cite{Crawford:2019} & Neuroblock & Scratch & User study & Introduce computer science concepts through Neuroblock \\
NeuroSquare~\cite{Mehul:2019} & Blockly & Support brain-computer interfaces using blocks and flow charts.\\
Neuroflow~\cite{Hernandez:2020} & Blockly & Block-flow environment for brain-computer interfaces. \\

% HCI experiments
Touchstone2~\cite{Eiselmayer:2020} & - & Tool to design HCI experiments. \\

\bottomrule
\end{tabular}
\caption{\Bb languages in human-computer interaction.}
\label{tab:hci}
\end{table}

%\subsection{Arts \& Creativity}
\begin{mybox2}[Arts \& Creativity]
\small
This category contains languages used for exploring creativity or as a medium for creating art through \bb constructs or by analyzing users' patterns as a result of their programming activities.
Languages that fall in this category are presented in \Cref{tab:creativity}.
\end{mybox2}

\begin{table}[htp]
\centering
\footnotesize
\begin{tabular}{l c l}
\toprule
{\textbf{Name}} & {\textbf{Metatool}} & {\textbf{Topic}}\\\midrule
% art
Quando~\cite{Stratton:2017} & Blockly & Create interactive digital exhibits for gallery visitors.\\
BlockArt\textsuperscript{\ref{exte}}~\cite{Dhariwal:2019} & Scratch & Visualize programming patterns in Scratch. \\

\bottomrule
\end{tabular}
\caption{\Bb languages in creativity.}
\label{tab:creativity}
%\vspace{-4mm}
\end{table}

%\subsection{Science}

\begin{mybox2}[Science]
\small
In this category, we found a single language using the block metaphor for conducting experiments in biology, see \Cref{tab:science}.
\end{mybox2}

\begin{table}[htp]
\centering
\footnotesize
\begin{tabular}{l c l}
\toprule
{\textbf{Name}} & {\textbf{Metatool}} & {\textbf{Topic}}\\\midrule
% biology
OpenLH~\cite{Gome:2019} & Blockly & Liquid handling system to conduct live biology experiments. \\
\bottomrule
\end{tabular}
\caption{\Bb languages in Science.}
\label{tab:science}
%\vspace{-4mm}
\end{table}

%\subsection{Artificial intelligence, data science, and databases}

\begin{mybox2}[Artificial intelligence, data science, and databases]
\small
This section contains \bb languages applied to the domain of artificial intelligence and data science.
This includes topics such as machine learning, chatbots, data science topics in general, and databases, as shown in \Cref{tab:ai}.
\end{mybox2}

\begin{table}[htp]
\footnotesize
\begin{tabular}{l c l}
\toprule
{\textbf{Name}} & {\textbf{Metatool}} & {\textbf{Topic}}\\\midrule

% ml
ScratchThAI\textsuperscript{\ref{exte}}~\cite{Katchapakirin:2018} & Scratch & Support computational thinking with a chatbot.\\
SnAIp\textsuperscript{\ref{exte}}~\cite{Jatzlau:20192} & Snap! & Enable machine learning within Snap!.\\
AlpacaML\textsuperscript{\ref{exte}}~\cite{Zimmermann:2020} & Scratch 3.0 & Test, evaluate, and refine ML models.\\

%data science
BlockPy~\cite{Bart:2017} & Blockly & Introductory programming environment for data science. \\
Scratch Community Blocks\textsuperscript{\ref{exte}}~\cite{Dasgupta:2017} & Scratch & Analysis and visualization of data coming from Scratch.\\
BlockArt\textsuperscript{\ref{exte}}~\cite{Dhariwal:2019} & Scratch & Visualization tool of programming in Scratch.\\
\cite{Fernandez:2020}\textsuperscript{\ref{exte}} & Scratch 3.0 & Engage learners in exploring and making sense of data.\\
Snap!DSS\textsuperscript{\ref{exte}}~\cite{Grillenberger:2017} & Snap! & Allow data stream analyses and visualization.\\

%DBs
DBSnap++~\cite{Silva:2018} & Snap! & Enable specification of dynamic data-driven programs. \\
DBSnap~\cite{Silva:2015} & - & Build database queries.\\
BlocklySQL~\cite{Nicolai:2019} & Blockly & \Bb editor for SQL.\\
DB-Learn~\cite{Vinayakumar:20181} & CT-Blocks & Teach relational algebra concepts.\\

\bottomrule
\end{tabular} 
\caption{\Bb languages in artificial intelligence and data science.}
\label{tab:ai}
%\vspace{-8mm}
\end{table}

%\subsection{Software Engineering}

\begin{mybox2}[Software engineering]
\small
This category contains different papers that present languages and proceedings that study \bbes usage in software engineering.
Therefore, the reader will find various topics such as code smells in \bb programs, security, testing, refactoring, debugging facilities, and specialized tools for developing \bb languages.
\end{mybox2}

In this category, we have grouped some papers that present a more theoretical view of the application of \bb languages.
Hermans and Aivaloglou~\cite{Hermans:2016} study code smells in the context of a \bbe, particularly in Scratch programs, and Techapalokul and Tilevich~\cite{Techapalokul:2017} study code quality in \bb programs using a methodology for code smells.
Swidan et al.~\cite{Swidan:2017} study naming patterns of Scratch programs' variables and procedures following this direction.
In contrast, Robles et al.~\cite{Robles:2017} identify software clones in Scratch projects.
The usage of static analysis techniques in \bb programs is beneficial, as shown by Jatzlau et al.~\cite{Jatzlau:2019}. They use static analysis techniques of Snap! programs to learn from programmers' behaviors.
Likewise, Aivaloglou and Hermans~\cite{Aivaloglou:2016} use static analysis techniques to explore Scratch programs' characteristics in software repositories.
Finally, Tenorio et al.~\cite{Tenorio:2019} study different debugging strategies in \bb programs.\\

% code smells- end-user programming
%\cite{Hermans:2016} studies code smells in the context of \bbe, particularly in Scratch programs
%\cite{Techapalokul:2017} studies code quality problems in \bb programs using a methodology of code smells.

%% naming
%\cite{Swidan:2017} study the naming patterns of variables and procedures in Scratch programs.
%
%% clones
%\cite{Robles:2017} software clones in scratch projects

% static analysis
%\cite{Jatzlau:2019} static analysis over Snap! programs to evaluate if the behavior is the same as in other \bb platforms.
%\cite{Aivaloglou:2016} static analysis of Scratch repository to explore characteristics of Scratch programs.

% debugging
%\cite{Tenorio:2019} studies the main debugging strategies in \bb programs.\\

\begin{table}[t]
\centering
\footnotesize
\begin{tabular}{lcl}
\toprule
{\textbf{Name}} & {\textbf{Metatool}} & {\textbf{Topic}}\\\midrule
ViSPE~\cite{Nergaard:2015} & Scratch & Policy editor for XACML.\\
XACML policy editor~\cite{Nergaard:20152} & Scratch & XACML policy editor.\\
\bottomrule
\end{tabular}
\caption{\Bb Languages in security.}
\label{tab:security}
%\vspace{-6mm}
\end{table}

\Cref{tab:security} shows two languages that we identified are used in topics related to security to define access control policies.
\Cref{tab:se} shows the list of languages used in different topics of software engineering.
Based on these tools, we highlight the appearance of one tool, \textit{Processing BBE}, designed specifically for creating \bbes.

\begin{table}[t]
\footnotesize
\centering
\begin{tabular}{lc m{8cm}}
\toprule
{\textbf{Name}} & {\textbf{Metatool}} & {\textbf{Topic}}\\\midrule

%testing
Extension Whisker~\cite{Stahlbauer:2019} & - & Testing framework for Scratch \\

%debugging
Extension~\cite{Savidis:2020} & Blockly & Add block-level debugging features to \bbe.\\
%block-based
Extension~\cite{Krutz:2019} & Blockly & Stepwise support for \bbes.\\

% TODO: tool for creating BBEs
Processing BBE~\cite{Kurihara:2015} & - & Create visual \bb domain-specific languages.\\

% Usability
Polymorphic Blocks~\cite{Lerner:2015} & - &  Represent complex structures and visual type information with \bb UI.\\

% bugs
LitterBox~\cite{Christoph:2020} & Scratch & Detecting bugs in Scratch programs.\\

%refactoring
QIS~\cite{Techapalokul:20192} & Scratch & Refactoring infrastructure for Scratch.\\
\cite{Techapalokul:2019}\textsuperscript{\ref{exte}} & Scratch 3.0 & Automated refactoring tool for Scratch.\\
%behavioral
Behavioral Blockly~\cite{ASHROV2015268} & Blockly & Support behavioral programming.\\

\bottomrule
\end{tabular}
\caption{\Bb Languages in software engineering.}
\label{tab:se}
%\vspace{-8mm}
\end{table}

\subsection{\Bb editors popularity}
So far, we have presented all the \bb languages that we identified in the papers included in this study.
As we have seen so far, most of the studies refer to Scratch as the most popular \bbe. 
To verify this, we manually kept track of the occurrences of each tool in each paper.
We started with an initial set of \bb languages that we obtained manually from searching at the most popular tools (see \Cref{sec:non-syst}).
When we had the initial set of languages, we proceeded to read the papers, and in a spreadsheet, we marked when a tool was mentioned and in which paper.
As we were reading papers, we added new languages that appeared to the set of \bb languages.
It is important to remark that in some cases, papers not only introduced a tool, but they also mention related tools that we also include in the list of tools.
This process has an explicit limitation since the discovery of languages is incremental as we read the papers.
Therefore, we made a sanity check using the same tool we developed and presented in \Cref{sec:qtwo} to mine the corpus of PDF files and collect the occurrences of each tool.

\begin{figure}[htp]
  \includegraphics[width=\textwidth]{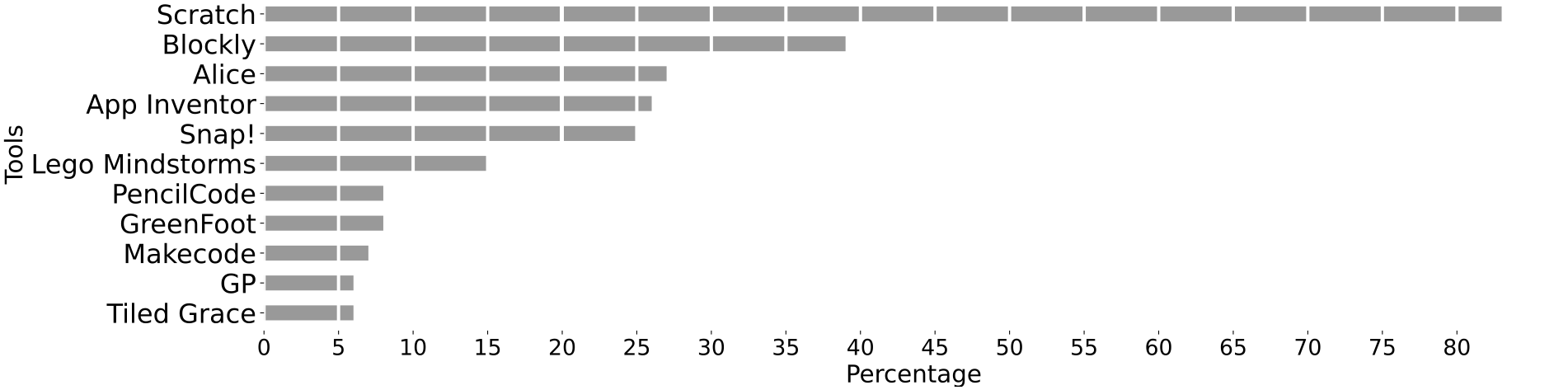}
  \caption{Popularity of \bbes across publications.}
  \label{fig:popularity_perc}
%  \vspace{-4mm}
\end{figure}

\Cref{fig:popularity_perc} shows a summary of the 11 most popular tools (see \Cref{ap:popularity} for the full list).
Since we used a program to mine the PDFs to double-check our manual results, the tool is not 100\% accurate.
In \Cref{sec:threats}, we present some of the limitations of the tool.

As speculated at the beginning, our results show that Scratch is indeed the language most mentioned in all the papers; it was mentioned in more than 80 of the papers of this study.
Similarly, Blockly is the second most mentioned language, even though it is not a language but a library for defining \bb languages.
The complete list of tools identified in this study and the number of papers in which they appear are shown in \Cref{ap:popularity}.

% TODO Thesis only

%This concludes the systematic part of our literature review.

\section{Non-Systematic Review of \bbes } \label{sec:non-syst}

As introduced before, we also conducted a less-systematic exploration using standard Google search to find information about \bbes\ that were not necessarily published academically.
\Cref{tab:industrial} presents a summary of our findings after analyzing and trying out each of the tools resulting from the search process.
Likewise, it also contains a set of features empirically collected by the first author after testing each language or tool.
The process to collect these features was by trying each tool and collecting its features in a spreadsheet.
Since all \bb environments do not offer the same features, a few tools had to be tested more than once because some features were included in the spreadsheet after testing the tool.
The table is divided into seven columns, and all columns except the first one are subdivided into other columns.

\begin{enumerate*}[label=(\roman*)]
	\item \textit{Name} represents the name of the tool or the language,
	\item \textit{Editor} represents the different components present in a code editor (e.g., mode, error marking, and stage),
	\item \textit{Focus} represents whether the tool is an application, a language that supports the developing \bb environments, or both,
	\item \textit{Deployment} shows the different models in which the tools are being offered, namely as standalone, mobile, or as Software as a Service (SaaS),
	\item \textit{Domain} represents the application domain where the tool is used,
	\item \textit{Execution} is how the tool executes an application. We identified mainly two modes: live and pressing an execution button (manual), and 
	\item \textit{Licensing} shows the three main types in which the tools are offered.
\end{enumerate*}

As the reader might have noticed, these features were used as a basis for the definition of the feature diagram of \bbes in \Cref{sec:qone}.
Using this manual exploration of all available tools we discovered most of the features of \bbes. The other features were discovered after the systematic literature review process described earlier. As described in the methodology, all tools listed in the table were tested by the first author.
Likewise, thanks to the mixed methodology, we were capable of identifying tools that we could not have discovered by relying solely on a systematic approach.
Therefore, the less-systematic exploration allowed us to discover 49 tools; from this number, only three tools (\textit{BlockPy}, \textit{CoBlox}, and \textit{Tuk tuk}) appeared in both the systematic and the less-systematic approach.
Some interesting facts of using this mixed methodology are discussed in \Cref{sec:disc}.

%\begin{landscape}

%\vspace*{\fill}

\begin{table}[t]
\centering
\footnotesize
\resizebox{\textwidth}{!}{%
\begin{tabular}{lllllllllllllllllllllllll}
\toprule
%\multicolumn{1}{c}{\multirow{3}{*}{\textbf{Name}}} & \multicolumn{7}{c}{\textbf{Editor}} & \multicolumn{2}{c}{\textbf{Focus}} & \multicolumn{3}{c}{\textbf{Deployment}} & \multicolumn{7}{c}{\textbf{Domain}} & \multicolumn{2}{c}{\textbf{Execution}} & \multicolumn{3}{c}{\textbf{Licensing}} \\
%\multicolumn{1}{c}{} & \multirow{2}{*}{\textbf{\begin{tabular}[c]{@{}c@{}}Single \\ mode\end{tabular}}} & \multirow{2}{*}{\textbf{\begin{tabular}[c]{@{}c@{}}Dual \\ mode\end{tabular}}} & \multicolumn{3}{c}{\textbf{Simulator}} & \multirow{2}{*}{\textbf{Feedback}} & \multirow{2}{*}{\textbf{\begin{tabular}[c]{@{}c@{}}Error \\ Marking\end{tabular}}} & \multirow{2}{*}{\textbf{Tool}} & \multirow{2}{*}{\textbf{Language}} & \multirow{2}{*}{\textbf{Standalone}} & \multirow{2}{*}{\textbf{Mobile}} & \multirow{2}{*}{\textbf{SaaS}} & \multirow{2}{*}{\textbf{Gaming}} & \multirow{2}{*}{\textbf{Animation}} & \multirow{2}{*}{\textbf{\begin{tabular}[c]{@{}c@{}}Programming \\ Hardware\end{tabular}}} & \multirow{2}{*}{\textbf{Programming}} & \multirow{2}{*}{\textbf{\begin{tabular}[c]{@{}c@{}}SE \\ Tools\end{tabular}}} & \multirow{2}{*}{\textbf{Education}} & \multirow{2}{*}{\textbf{\begin{tabular}[c]{@{}c@{}}multi-agent \\ modeling \\ environment\end{tabular}}} & \multirow{2}{*}{\textbf{Live}} & \multirow{2}{*}{\textbf{Manual}} & \multirow{2}{*}{\textbf{Academic}} & \multirow{2}{*}{\textbf{Commercial}} & \multirow{2}{*}{\textbf{Open-source}} \\
%\multicolumn{1}{c}{} &  &  & \textbf{Textual} & \textbf{Visual} & \textbf{Hardware} &  &  &  &  &  &  &  &  &  &  &  &  &  &  &  &  &  &  &  \\\midrule
 & \multicolumn{7}{c}{\textbf{Editor}} & \multicolumn{2}{c}{\textbf{Focus}} & \multicolumn{3}{c}{\textbf{Deployment}} & \multicolumn{7}{c}{\textbf{Domain}} & \multicolumn{2}{c}{\textbf{Execution}} & \multicolumn{3}{c}{\textbf{Licensing}} \\  \cmidrule(l){2-8}  \cmidrule(l){9-10}  \cmidrule(l){11-13}  \cmidrule(l){14-20}  \cmidrule(l){21-22}  \cmidrule(l){23-25}
 &  &  & \multicolumn{3}{c}{\textbf{Stage}} &  &  &  &  &  &  &  &  &  &  &  &  &  &  &  &  &  &  & \\ \cline{4-6}
\multicolumn{1}{c}{\multirow{1}{*}{\textbf{Name}}} & \rot[90]{\textbf{Single mode}} & \rot[90]{\multirow{2}{*}{\textbf{Dual mode}}} & \rot[90]{\multirow{2}{*}{\textbf{Textual}}} & \rot[90]{\multirow{2}{*}{\textbf{Visual}}} & \rot[90]{\multirow{2}{*}{\textbf{Hardware}}} & \rot[90]{\multirow{2}{*}{\textbf{Feedback}}} & \rot[90]{\multirow{2}{*}{\textbf{Error marking}}} & \rot[90]{\multirow{2}{*}{\textbf{Application}}} & \rot[90]{\multirow{2}{*}{\textbf{Language}}} & \rot[90]{\multirow{2}{*}{\textbf{Standalone}}} & \rot[90]{\multirow{2}{*}{\textbf{Mobile}}} & \rot[90]{\multirow{2}{*}{\textbf{SaaS}}} & \rot[90]{\multirow{2}{*}{\textbf{Gaming}}} & \rot[90]{\multirow{2}{*}{\textbf{Animation}}} & \rot[90]{\multirow{2}{*}{\textbf{\begin{tabular}[c]{@{}c@{}}Programming \\ Hardware\end{tabular}}}} & \rot[90]{\multirow{2}{*}{\textbf{Programming}}} & \rot[90]{\multirow{2}{*}{\textbf{SE tools}}} & \rot[90]{\multirow{2}{*}{\textbf{Education}}} & \rot[90]{\multirow{2}{*}{\textbf{\begin{tabular}[c]{@{}c@{}}Multi-agent \\ modeling\end{tabular}}}} & \rot[90]{\multirow{2}{*}{\textbf{Live}}} & \rot[90]{\multirow{2}{*}{\textbf{Manual}}} & \rot[90]{\multirow{2}{*}{\textbf{Academic}}} & \rot[90]{\multirow{2}{*}{\textbf{Commercial}}} & \rot[90]{\multirow{2}{*}{\textbf{Open-source}}} \\ \hline
Pencil code &  & $\sbullet$ & $\sbullet$ & $\sbullet$ &  & $\sbullet$ &  & $\sbullet$ &  &  &  & $\sbullet$ &  & $\sbullet$ &  & $\sbullet$ &  & $\sbullet$ &  &  & $\sbullet$ & $\sbullet$ &  & $\sbullet$ \\
Scratch & $\sbullet$ &  &  & $\sbullet$ &  &  &  & $\sbullet$ &  &  &  & $\sbullet$ & $\sbullet$ & $\sbullet$ &  &  &  & $\sbullet$ &  &  & $\sbullet$ &  &  & $\sbullet$ \\
Alice & $\sbullet$ &  &  & $\sbullet$ &  &  &  & $\sbullet$ &  & $\sbullet$ &  &  & $\sbullet$ & $\sbullet$ &  &  &  & $\sbullet$ &  &  & $\sbullet$ & $\sbullet$ &  &  \\
Blockly & $\sbullet$ &  &  &  &  &  &  &  & $\sbullet$ & $\sbullet$ & $\sbullet$ & $\sbullet$ & $\sbullet$ & $\sbullet$ & $\sbullet$ & $\sbullet$ & $\sbullet$ & $\sbullet$ &  &  &  &  &  & $\sbullet$ \\
Mindstorms & $\sbullet$ &  &  &  & $\sbullet$ &  &  & $\sbullet$ &  & $\sbullet$ &  &  &  &  & $\sbullet$ &  &  & $\sbullet$ &  &  & $\sbullet$ &  & $\sbullet$ &  \\
App inventor &  & $\sbullet$ &  & $\sbullet$ & $\sbullet$ & $\sbullet$ & $\sbullet$ & $\sbullet$ &  & $\sbullet$ &  & $\sbullet$ &  &  & $\sbullet$ & $\sbullet$ &  & $\sbullet$ &  &  & $\sbullet$ & $\sbullet$ &  & $\sbullet$ \\
Pocket Code & $\sbullet$ &  &  &  &  &  &  & $\sbullet$ &  &  & $\sbullet$ &  & $\sbullet$ & $\sbullet$ &  &  &  & $\sbullet$ &  &  & $\sbullet$ & $\sbullet$ &  &  \\
Deltatick &  & $\sbullet$ & $\sbullet$ &  &  &  &  & $\sbullet$ &  & $\sbullet$ &  &  &  &  &  &  &  & $\sbullet$ & $\sbullet$ &  & $\sbullet$ & $\sbullet$ &  &  \\
Frog pond & $\sbullet$ &  &  & $\sbullet$ &  & $\sbullet$ &  & $\sbullet$ &  &  &  & $\sbullet$ &  & $\sbullet$ &  &  &  & $\sbullet$ &  &  & $\sbullet$ & $\sbullet$ &  &  \\
StarLogo TNG & $\sbullet$ &  &  & $\sbullet$ &  &  &  & $\sbullet$ &  & $\sbullet$ &  &  & $\sbullet$ &  &  & $\sbullet$ &  & $\sbullet$ &  &  &  & $\sbullet$ &  &  \\
Turtle art & $\sbullet$ &  &  & $\sbullet$ &  &  &  & $\sbullet$ &  & $\sbullet$ &  &  &  & $\sbullet$ &  &  &  & $\sbullet$ &  &  & $\sbullet$ & $\sbullet$ &  &  \\
PicoBlocks & $\sbullet$ &  &  &  &  &  &  & $\sbullet$ &  & $\sbullet$ &  &  &  &  & $\sbullet$ &  &  & $\sbullet$ &  &  & $\sbullet$ &  & $\sbullet$ &  \\
Robobuilder & $\sbullet$ &  &  & $\sbullet$ &  &  &  & $\sbullet$ &  & $\sbullet$ &  &  & $\sbullet$ & $\sbullet$ &  &  &  & $\sbullet$ &  &  & $\sbullet$ & $\sbullet$ &  &  \\
OpenBlocks &  &  &  &  &  &  &  &  & $\sbullet$ &  &  &  &  &  &  &  &  &  &  &  &  &  &  & $\sbullet$ \\
miniBloq & $\sbullet$ &  &  &  & $\sbullet$ &  &  & $\sbullet$ &  & $\sbullet$ &  &  &  &  & $\sbullet$ &  &  & $\sbullet$ &  &  & $\sbullet$ &  &  & $\sbullet$ \\
GP & $\sbullet$ &  &  & $\sbullet$ &  &  &  & $\sbullet$ &  & $\sbullet$ &  & $\sbullet$ &  & $\sbullet$ &  & $\sbullet$ &  & $\sbullet$ &  &  & $\sbullet$ &  &  &  \\
Snap! & $\sbullet$ &  &  & $\sbullet$ &  &  &  & $\sbullet$ &  &  &  & $\sbullet$ &  & $\sbullet$ &  & $\sbullet$ &  & $\sbullet$ &  &  & $\sbullet$ &  &  & $\sbullet$ \\
Microblocks & $\sbullet$ &  &  & $\sbullet$ & $\sbullet$ &  &  & $\sbullet$ &  & $\sbullet$ &  &  &  &  & $\sbullet$ & $\sbullet$ &  & $\sbullet$ &  & $\sbullet$ &  &  &  & $\sbullet$ \\
Makecode & $\sbullet$ & $\sbullet$ & $\sbullet$ &  &  &  &  & $\sbullet$ & $\sbullet$ & $\sbullet$ &  & $\sbullet$ & $\sbullet$ &  & $\sbullet$ & $\sbullet$ &  & $\sbullet$ &  & $\sbullet$ & $\sbullet$ &  &  &  \\
Waterbear & $\sbullet$ &  &  & $\sbullet$ &  &  &  & $\sbullet$ &  &  &  & $\sbullet$ &  & $\sbullet$ &  & $\sbullet$ &  & $\sbullet$ &  &  & $\sbullet$ &  &  & $\sbullet$ \\
Looking Glass & $\sbullet$ &  &  & $\sbullet$ &  &  &  & $\sbullet$ &  & $\sbullet$ &  &  & $\sbullet$ & $\sbullet$ &  &  &  & $\sbullet$ &  &  & $\sbullet$ & $\sbullet$ &  &  \\
%Greenfoot’s frame- based Stride editor
Greenfoot &  & $\sbullet$ &  & $\sbullet$ &  &  &  & $\sbullet$ &  & $\sbullet$ &  &  & $\sbullet$ &  &  & $\sbullet$ &  & $\sbullet$ &  &  &  & $\sbullet$ &  &  \\
Applab &  & $\sbullet$ & $\sbullet$ & $\sbullet$ &  & $\sbullet$ & $\sbullet$ & $\sbullet$ &  &  &  & $\sbullet$ &  &  &  & $\sbullet$ &  & $\sbullet$ &  &  & $\sbullet$ &  &  & $\sbullet$ \\
StarLogo Nova &  &  & $\sbullet$ &  &  &  &  & $\sbullet$ &  &  &  &  &  &  &  &  &  &  &  &  &  &  &  &  \\
Tynker &  &  & $\sbullet$ & $\sbullet$ &  &  &  & $\sbullet$ &  &  &  &  &  &  &  &  &  &  &  &  &  &  &  &  \\
Hopscotch &  &  &  &  &  &  &  & $\sbullet$ &  &  &  &  &  &  &  &  &  &  &  &  &  &  &  &  \\
AutoBlocks & $\sbullet$ &  &  &  &  &  &  & $\sbullet$ &  & $\sbullet$ &  &  &  &  &  &  &  &  &  &  &  &  &  &  \\
BlockPy & $\sbullet$ & $\sbullet$ & $\sbullet$ &  &  & $\sbullet$ & $\sbullet$ & $\sbullet$ &  &  &  & $\sbullet$ &  &  &  & $\sbullet$ &  & $\sbullet$ &  &  & $\sbullet$ & $\sbullet$ &  & $\sbullet$ \\
Kodika &  & $\sbullet$ &  &  & $\sbullet$ &  &  & $\sbullet$ &  &  & $\sbullet$ &  &  &  & $\sbullet$ & $\sbullet$ &  &  &  &  &  &  &  &  \\
Sphero Sprk & $\sbullet$ &  &  &  & $\sbullet$ &  &  & $\sbullet$ &  & $\sbullet$ & $\sbullet$ & $\sbullet$ & $\sbullet$ &  & $\sbullet$ & $\sbullet$ &  & $\sbullet$ &  &  & $\sbullet$ &  & $\sbullet$ &  \\
Stencyl &  & $\sbullet$ &  & $\sbullet$ & $\sbullet$ & $\sbullet$ & $\sbullet$ &  &  & $\sbullet$ &  &  & $\sbullet$ &  &  & $\sbullet$ &  & $\sbullet$ &  &  &  &  & $\sbullet$ &  \\
Tynker &  &  & $\sbullet$ & $\sbullet$ &  &  &  & $\sbullet$ &  &  &  &  &  &  &  &  &  &  &  &  &  &  &  & \\\bottomrule
\end{tabular}%
}
\caption{Tools identified using the non systematic approach via standard Google search.}
\label{tab:industrial}
%\vspace{-8mm}

\end{table}
%\vspace*{\fill}
%
%\end{landscape}

%\todo{list systems not found by the structured review, and list features not found by the structured literature review}
%\todo{explain how you collected the features. in other parts you explain stuff about spreadsheets and stuff}

\section{Threats to Validity}
\label{sec:threats}
A systematic literature review (SLR) is a research methodology used to obtain a complete overview of a particular topic or domain.
Based on that, we followed the Kitchenham et al.~\cite{kitchenham:2009} guidelines, and we defined our protocol for conducting this study. We identified some threats to validity that we discuss in more detail in this section.

\subsection{External validity}
SLRs are conducted to present a summary of a particular topic or domain.
Although authors try to reduce their bias as much as possible, it is almost impossible to eradicate it. 
Thus, this is one of the main threats to validity and a critical aspect of these studies. 
In the design of our protocol, we tried to minimize as much as possible our bias by defining three filters for including the final set of papers.
Moreover, two authors discussed the inclusion and exclusion criteria for a sample of ten papers.
Nonetheless, it is essential to mention that since we were looking at specific research questions, this study can never be entirely unbiased, and it is focused on addressing these questions.
The queries and the sources of information used in this study prevent us from being fully unbiased.
Nonetheless, we tried to keep the current study as broad as possible; in the paper selection, a wide variety of papers came from different communities, venues, and areas of expertise.
Moreover, in general, the notion of \bbes is ambiguous; this term is used to refer to two different topics.
On the one hand, visual programming environments that adopt the jigsaw metaphor for creating programs (discussed in this paper), and on the other hand, the notion of blocks in a block diagram (e.g., Simulink), which is often used in simulation applications and model-based design.

\subsection{Internal validity}
% tools
Since the data collection was a manual task, we consider it essential to conduct a sanity check using automated tools.
For this purpose, we developed a tool for scanning and mining PDF files and checks whether a given list of words appears in the file's content.
There are some known caveats which concern the accuracy and correctness of the tool.
First, reading and mining PDF files is not an easy task, mainly because PDF files do not share a standard structure.
Thus, some files cannot be opened, or all the text is not parseable.
Second, the list of words was manually defined.
In the case of programming languages popularity, it was obtained from the TIOBE index~\cite{tiobe:2021}, which made it more accessible.
However, to double-check the languages' popularity, this list was a manual process, which started from a list of languages obtained via a non-systematic method.
This list of languages was improved by taking manual notes of new tools presented in papers and their related work.
Therefore, it could be the case that the last paper read by the authors introduced a new tool, which of course, was not marked in the previous papers since it was not found yet.
However, thanks to the automated tool, we can detect across all the papers if the tool is mentioned or not.
In this direction, the tool results are not 100 \% accurate due to different factors.
\begin{enumerate*}[label=(\roman*)]
	\item \textit{Ambiguous words}. Words in the input list are valid words in English. For instance, \textit{Scratch} or \textit{Go}.
Thus, the tool does not differentiate whether it is an English word or refers to a \bb editor or a programming language.
	\item \textit{Punctuation marks}. The tool compares word by word each of the words in the input list against the text. This means that if a word in the input text appears in the text next to a punctuation mark (e.g., colon or comma), the tool produces a false negative result. The tool says that the word is not present, even though it is present, but it does not capture it since it is next to a punctuation mark (without a blank space in the middle).
\end{enumerate*}
 To measure possible errors in the tool, we sampled ten papers and five programming languages to check how accurate the tool's results are.
We calculated type I and type II errors based on the sample to identify the numbers of false positives and false negatives, respectively.
The results obtained show that the sensitivity of the tool is 75\%. 
This means that there is a rate of false negatives of 25\%.
In other words, in 25\% of the cases, the tool says that the word is not present in the document, but it is.
Similarly, the tool's specificity is about 82,6\%, which means that the false-positive rate is 17,4\%.
In 17\% of the cases, the tool said a word was present in the document, even though the word was not present.

In both cases, the tool can be fine-tuned so that both the sensitivity and specificity improve by considering the corner cases previously mentioned.
However, that is not the main focus of the current paper. 
We developed this tool as a sanity check to refine the results obtained during the manual inspection.
%\subsection{Tool}
% PLS

In \Cref{sec:qtwo}, where we present the programming languages used, some papers do not mention how they were implemented. For instance, we could have assumed that when they use Blockly, the editor was implemented using JavaScript, which is the most popular language used for using Blockly.
Nevertheless, this is not true for all the cases, because it is also available in other programming languages.
Therefore, we decided not to make assumptions about this.

As presented in the protocol, we only considered four academic databases to obtain the academic papers, and the non-systematic search gave us practical languages that do not necessarily have an academic publication.
However, the latter means that this part is not easily reproducible.

\section{Discussion} \label{sec:disc}
We identified three main ways that developers follow to create \bbes.
% 1st and 2nd are similar, maybe the extensions are just a subgroup
The first approach is by extending an existing language. 
Twenty-seven of the languages included in this study were developed using this approach.
The second one is by using a library that supports the development of such languages.
As expected, this is the most popular solution we found in the tools we discovered. 
Sixty-one languages were developed using other libraries since this reduces the development effort.
Finally, the third option is a bespoke implementation.
Based on our corpus, only nine languages used this approach.
It is important to remark that the previous methods for implementing \bbes are defined based only on our observations.
This might not be true for all cases, given that many authors did not mention any implementation details.

%extension: 27
%ad-hoc: 9
%metatool: 88

It is interesting to see in the data that there are not many tools that support the whole development cycle of \bbes.
There are specialized libraries for creating concrete pieces of them, but most of these environments rely on code generators. 
For instance, Blockly is used for describing the UI of the language, and then programs must be compiled to a target language (e.g., Python).
We found two tools (\cite{Kurihara:2015,Verano:2020}) to develop software languages with a built-in \bb editor. 
However, these two tools are relatively new or not widely adopted; none of the languages presented in this paper was implemented using them.
Likewise, it is relevant to mention that the approach proposed by ~\cite{Kurihara:2015} relies on code generators.
Instead, \cite{Verano:2020} relies on language workbench technology for defining both the syntax and the semantics of languages, which makes such languages also usable outside a \bb editor in a traditional IDE.

Based on the collected data, it is evident that the most popular programming language for implementing \bbes is JavaScript (\Cref{tab:meta-tools}).
This seems an interesting outlier, but it should not be seen independently from the following observation.
Most \bbes were implemented using Blockly, which is a library implemented in JavaScript. 
Even though, Blockly offers implementations in other programming languages (e.g., Swift), these have been deprecated and are no longer maintained by the Blockly Team.
Moreover, several \bb languages are implemented as web applications, which also explains the vast popularity of using JavaScript for creating \bb languages.

As shown in \Cref{sec:qthree}, there is a limited number of libraries for developing \bbes.
Therefore, we see that many authors rely on existing \bbes to build their own.
Surprisingly, specialized language engineering tools (e.g., LWBs) are not used in this domain.
JastAdd~\cite{jastadd} and ANTLR~\cite{antlr} were used for developing two environments, each one.
Our research resulted in \textit{Kogi}~\cite{Verano:2020}, that uses the Rascal LWB~\cite{rascalscam} to create \bb editors for new and existing languages. 
This to make \bb editors part of the generic services offered by LWBs. 
However, this tool was not considered in this survey because it was published afterwards.

Another interesting observation that resulted from this study is using mixed methods (systematic and non-systematic searches).
As presented in this survey, we see differences between the results obtained from the systematic literature review and the non-systematic tool review.
We identified some hypotheses behind these differences.
First, some tools are developed to address a specific problem, which is not always followed by a scientific publication.
Moreover, there are also industrial applications. 
Their primary focus is not necessarily the development of scientific publications and following existing literature but to address business requirements and make things work.
Another critical aspect of industrial applications is their visibility; sometimes, they are not disclosed due to intellectual property rights.
As we underlined in our data, the difference is remarkable. 
From the 35 languages and tools that we identified in the non-systematic approach, only 3 had a research paper included in this review.
This means that more than 91\% of the tools would not have been included if we did not conduct a search of non-academic literature and tools.

%\section{What are the benefits drawbacks of using \bb}

%  Use google's Natural Language API to make a sentiment analysis on the extracted data from the reading.

\section{Related work}\label{sec:related}
Coronado et al.~\cite{CORONADO:2020} present a literature review about 16 visual programming environments to foster end-user development (EUD) of applications that involve robots with social capabilities.
This survey focuses on visual programming environments for non-professional programmers, and they highlight mainly two goals.
The first one is to present a list of the tools with their technical features, and the second, to present the open challenges in the development of visual programming environments for end-users. 
McGill and Decker~\cite{mcgill:2020} conducted a systematic literature review and propose a taxonomy for tools, languages, and environments (TLEs) used in computer education.
Their main focus is on studying TLEs used in primary, secondary, and post-secondary education.
Based on their study, they propose a TLEs taxonomy.
Solomon et al.~\cite{Solomon:2020} present the history of Logo, a programming environment designed for children to explore mathematical concepts.
This is the main predecessor of current notions of \bbes for end-users.

Rough and Quigley~\cite{Rough:2020} present a perspective of end-user development (EUD) for creating and customizing software by end-users, as end-users outnumbered professional programmers.
As a result of their work, they propose some design recommendations to support EUD activities, particularly the creation of software that allows novice users to create apps that collect data (e.g., experience sampling).
This paper follows a similar methodology. They queried computer science databases and a non-systematic approach through Google search to get non-academic tools.
%Section 2.2. \cite{Rough:2020} described the same kind of methodology I followed.

\section{Conclusions and Future Work}\label{sec:conclusions}
This paper presents an overview of \bbes and their features. 
Also, it presents a detailed view of how these programming environments are developed and the technologies involved in this process.
We listed and summarized more than one hundred languages and extensions, which were grouped into seven categories.
These categories highlight the fact that \bbes have a broader scope than computer science education.
The results show that authors do not mention implementation details or possible troubles that the development of a \bb editor has.
Moreover, there is a vast diversity of applications in which the \bb metaphor is adopted (e.g., arts, education, science, robotics). Yet, there is a lack of tool support for developing a whole language that supports a \bb editor. Existing tools do not support the whole development cycle of a language. 
In most cases, designers of \bbes rely on code generators for defining the semantics of the languages.
We believe that the usage of meta-programming technologies, such as found in Language Workbenches, would enable engineers to fully develop a language and obtain a \bb editor almost ``for free'', as is the case already for textual editors.
Likewise, we confirmed that Scratch has had a significant impact on the development of most of current \bbes, both conceptually and technically.

Another interesting conclusion of the current survey is that using different methods and sources (systematic and less-systematic, academic and non-academic) allowed us to synthesize a more complete overview of this particular topic than would otherwise be possible.
In particular, the less-systematic approach to collect information from non-academic sources presented findings complementary to the systematic literature study, which were also fundamental to the interpretation of the data from the systematic literature study.

We also provided an overview of academic research on usability and learnability of block-based editors (as compared to text editors) and other studies of large collections of block-based programs.

As future work, we foresee different directions:
\begin{enumerate*}[label=(\roman*)]
	\item Study what are the best practices for using and implementing \bb editors. The current paper presents an overview of the features we identified across languages. However, it is interesting to explore the particularities of \bb interfaces to improve the users' programming experience; and how this can be used to implement better \bb editors.
	\item Explore the integration of \bb editors as part of the default set of services offered by specialized tooling for language development (e.g., language workbenches).
	\item Study the lack of tools that support the creation and generation of \bb editors.
	\item Support dynamic aspects of languages in \bbes (e.g., debugging and live programming).
	\item Investigate hybrid environments in which parts of blocks are text-based, depending on the language construct's nature. This to improve the usability and efficiency between drag and drop and writing textual code.
\end{enumerate*}

%Open challenges for \bb editors.

%Best practices for using and implementing \bb editors.

%investigate difficulties for implementing them

%Make \bb editors part of the default set of services offered by language workbenches.

% lack of tools
%This is interesting, and worth to study in the future, perhaps there is a lack of specialized tools for building \bbes?

%\nocite{*}

%\begin{acks}
%To Robert, for the bagels and explaining CMYK and color spaces.
%\end{acks}

\bibliographystyle{ACM-Reference-Format}
\bibliography{bibliography}

\newpage
\appendix

\section{Phase 2: filtering questions}
\label{ap:questions}

\begin{itemize}
	\item Is the publication a full paper?
	\item Does the paper introduce a language or a tool that uses a \bb editor?
	\item Does the paper introduce a tool for building \bbes?
	\item Does the paper use or study \bbes?
	\item Does the paper present implementation details regarding the \bbe?
	\item Does the paper present best practices for using \bbes?
	\item Does the paper present best practices or guidelines for implementing \bbes?
	\item Does the paper present limitations of \bbes?
	\item Does the paper present open challenges that should be addressed with \bbes?
\end{itemize}

\section{Programming languages Popularity}
\label{ap:pls}

\begin{table}[htp]
\centering	
\footnotesize
\begin{tabular}{lc}
\toprule
     \textbf{Language} & \textbf{Occurrences} \\
\midrule
      Scratch &          124 \\
            C &           52 \\
         Java &           50 \\
           Go &           46 \\
            R &           44 \\
   JavaScript &           44 \\
            D &           39 \\
       Python &           35 \\
         Logo &           28 \\
       Scheme &           16 \\
          C++ &           15 \\
          SQL &            9 \\
          NaN &            8 \\
          PHP &            6 \\
        Julia &            6 \\
      Haskell &            5 \\
           C\# &            3 \\
        Swift &            3 \\
   TypeScript &            3 \\
      LabVIEW &            2 \\
        Scala &            2 \\
       MATLAB &            1 \\
           PL &            1 \\
       Prolog &            1 \\
 Visual Basic &            1 \\
         Perl &            1 \\
       Groovy &            1 \\
          Lua &            1 \\
          SAS &            1 \\
          Ada &            1 \\
\bottomrule
\end{tabular}
\caption{pls popularity}
\end{table}

\newpage
\section{Papers per Venue}

\begingroup
\footnotesize
\begin{longtable}{m{2.2cm}m{30em}c}
\toprule
\textbf{Venue} & \textbf{Category} & \textbf{\# Papers} \\
\midrule
\endhead
\midrule
\multicolumn{2}{r}{{Continued on next page}} \\
\midrule
\endfoot
\multirow{12}{*}{\begin{tabular}[c]{@{}l@{}}Human computer\\ interaction\end{tabular}} & ACM Conference on Human Factors in Computing Systems (CHI) & 13 \\
 & ACM Conference on Interaction Design and Children & 11 \\
 & International Conference on Human-Computer Interaction (HCII) & 4 \\
 & International Journal of Child-Computer Interaction & 3 \\
 & Creativity and Cognition & 1 \\
 & International Conference on Tangible, Embedded, and Embodied Interaction & 1 \\
 & IFIP Conference on Human-Computer Interaction (INTERACT) & 1 \\
 & ACM Symposium on User Interface Software and Technology (UIST) & 1 \\
 & International Conference of Design, User Experience, and Usability (DUXU) & 1 \\
 & International Symposium on End User Development & 1 \\
 & Iberoamerican Workshop on Human-Computer Interaction (HCI-COLLAB) & 1 \\
 & \begin{tabular}[c]{@{}l@{}}International Conference on Human Systems \\ Engineering and Design: Future Trends and Applications (IHSED)\end{tabular} & 1 \\
\midrule
\multirow{4}{*}{\begin{tabular}[c]{@{}l@{}}Programming /\\ Human computer \\ Interaction\end{tabular}} & IEEE Blocks and Beyond Workshop (Blocks and Beyond) & 13 \\
 & IEEE Symposium on Visual Languages and Human-Centric Computing (VL/HCC) & 9 \\
 & Journal of Visual Languages \& Computing & 2 \\
 & International Symposium on End User Development (IS-EUD) & 2 \\
\midrule
\multirow{25}{*}{Education} & ACM Technical Symposium on Computer Science Education (SIGSE) & 10 \\
 & IEEE Global Engineering Education Conference (EDUCON) & 5 \\
 & Computational Thinking Education & 2 \\
 & Education and Information Technologies & 2 \\
 & Workshop in Primary and Secondary Computing Education & 2 \\
 & International Conference on International Computing Education Research & 2 \\
 & ACM Conference on International Computing Education Research & 2 \\
 & ACM Transactions on Computing Education & 1 \\
 & International Conference on Information Technology Based Higher Education and Training (ITHET) & 1 \\
 & International Conference on Learning and Teaching in Computing and Engineering (LaTICE) & 1 \\
 & Journal of Computing Sciences in Colleges & 1 \\
 & Journal of computer sciences in colleges & 1 \\
 & Innovation and Technology in Computer Science Education (ITiCSE) & 1 \\
 & Workshop in Primary and Secondary Computing Education (WIPSCE) & 1 \\
 & ACM Conference on Innovation and Technology in Computer Science Education & 1 \\
 & International Journal of Artificial Intelligence in Education & 1 \\
 & International Conference on Informatics in Schools: Situation, Evolution, and Perspectives (ISSEP) & 1 \\
 & Computers \& Education & 1 \\
 & International Conference on Artificial Intelligence in Education (AIED) & 1 \\
 & Journal of Science Education and Technology & 1 \\
 & International Conference on Blended Learning (ICBL) & 1 \\
 & International Conference on Computer Supported Education (CSEDU) & 1 \\
 & International Conference on Interactive Collaborative Learning (ICL) & 1 \\
 & International Conference on Learning and Collaboration Technologies (LCT) & 1 \\
 & Software Data Engineering for Network eLearning Environments & 1 \\
\midrule
\multirow{5}{*}{Distributed computing} & International Conference on Computing, Communication and Networking Technologies (ICCCNT) & 5 \\
 & Proceedings of the ACM on Interactive, Mobile, Wearable and Ubiquitous Technologies & 1 \\
 & IEEE International Smart Cities Conference (ISC2) & 1 \\
 & IEEE International Parallel and Distributed Processing Symposium Workshops (IPDPSW) & 1 \\
 & Journal of Parallel and Distributed Computing & 1 \\
\midrule
\multirow{2}{*}{Robotics / Education} & International Conference on Robotics and Education (RiE) & 4 \\
 & Robotics in Education & 1 \\
 \midrule
\begin{tabular}[c]{@{}l@{}}Accessibility/\\  Human computer Interaction\end{tabular} & ACM SIGACCESS Conference on Computers and Accessibility (ASSETS) & 2 \\
\midrule
\multirow{2}{*}{Programming} & Science of Computer Programming & 2 \\
 & \begin{tabular}[c]{@{}l@{}}ACM SIGPLAN International Conference on Systems, Programming, \\ Languages, and Applications: Software for Humanity (Splash-E)\end{tabular} & 1 \\
\midrule
Security & International Conference on Information Systems Security and Privacy (ICISSP) & 2 \\
\midrule
\multirow{9}{*}{Software engineering} & IEEE International Working Conference on Source Code Analysis and Manipulation (SCAM) & 2 \\
 & IEEE International Conference on Program Comprehension (ICPC) & 1 \\
 & \begin{tabular}[c]{@{}l@{}}ACM Joint Meeting on European Software Engineering Conference \\ and Symposium on the Foundations of Software Engineering\end{tabular} & 1 \\
 & Conference of the Center for Advanced Studies on Collaborative Research (CASCON) & 1 \\
 & IEEE International Workshop on Software Clones (IWSC) & 1 \\
 & IEEE Transactions on Emerging Topics in Computing & 1 \\
 & International Conference on Soft Computing and Software Engineering (SCSE'15) & 1 \\
 & Journal of Systems Architecture & 1 \\
 & IEEE Conference on Open Systems (ICOS) & 1 \\
 \midrule
Technology & International Conference on Advances in Information Technology & 1 \\
\midrule
\multirow{4}{*}{General} & Communications of the ACM & 1 \\
 & IEEE Access & 1 \\
 & TechTrends & 1 \\
 & Science and Information Conference (SAI) & 1 \\
 \midrule
\multirow{2}{*}{Engineering} & International Conference on Automation, Computational and Technology Management (ICACTM) & 1 \\
 & International Conference on Developments in eSystems Engineering (DeSE) & 1 \\
 \midrule
\multirow{2}{*}{Robotics} & Iberian Robotics conference (Robot) & 1 \\
 & International Conference on Ubiquitous Robots (UR) & 1 \\
AI & ITS (Intelligent Tutoring Systems) 2020 & 1 \\
\midrule
eBusiness & International Conference on E-Business and Applications (ICEBA) & 1 \\
\midrule
Games & International Conference on Serious Games, Interaction, and Simulation (SGAMES) & 1 \\
\midrule
Electrical engineering & Computers \& Electrical Engineering & 1 \\
\midrule
Multimedia & Multimedia Tools and Applications & 1\\
\end{longtable}
\label{ap:venues}
\endgroup

%%%%%%%%%%%%%%%

\newpage
\section{Papers per country}
\label{ap:countries}

\begin{figure}[H]
              \includegraphics[width=\linewidth]{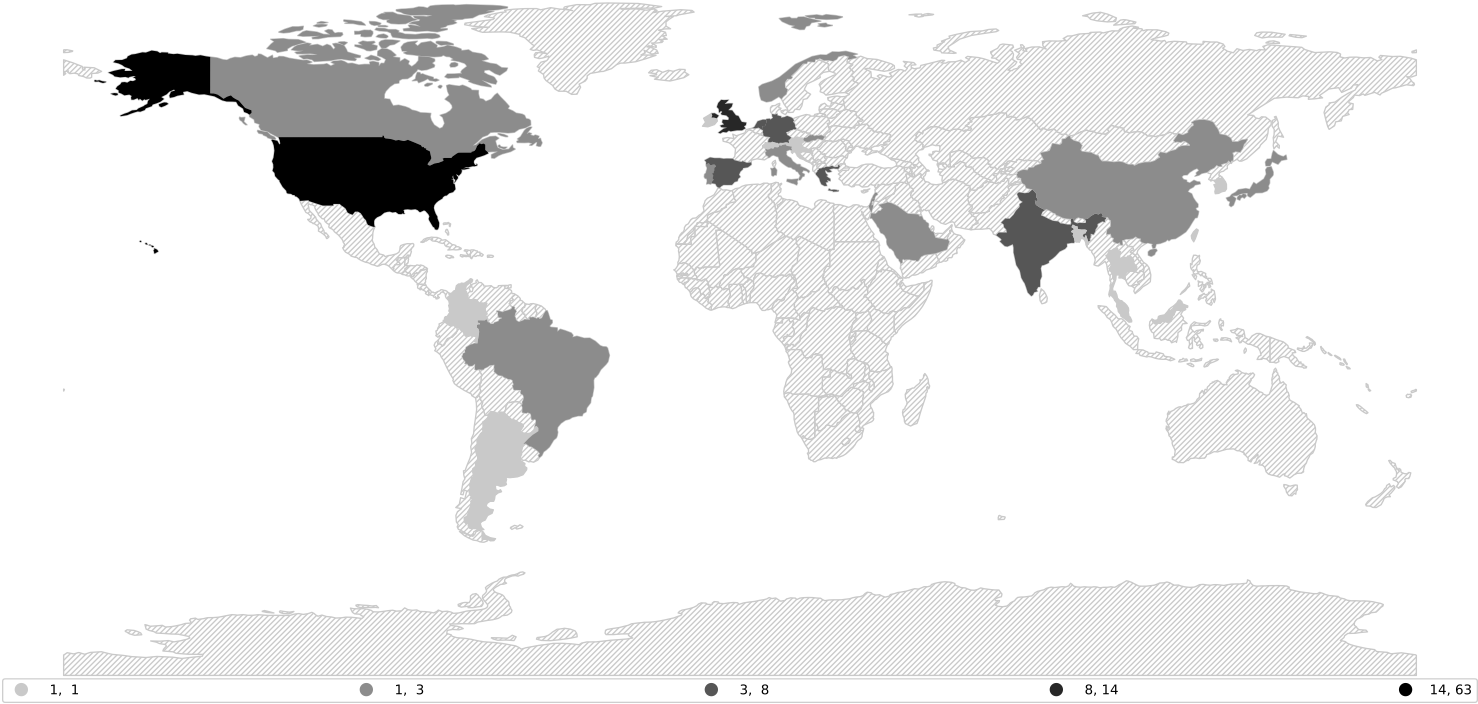}
              \caption{This choropleth map displays a summary of the number of papers published per country. The detailed table with the values is shown in \Cref{tb:country}.}
              \label{fig:papers_per_country}
          \end{figure}

\begin{table}[htp]
\footnotesize
\centering
\begin{tabular}{lc}
\toprule
\textbf{Country} &  \textbf{Papers} \\
\midrule
          USA &      64 \\
           UK &      14 \\
      Germany &       8 \\
        Spain &       6 \\
        India &       6 \\
       Greece &       5 \\
  Netherlands &       5 \\
   New Zealand &       4 \\
     Portugal &       3 \\
       Brazil &       3 \\
       Norway &       3 \\
        Japan &       3 \\
       Israel &       3 \\
       Canada &       2 \\
        China &       2 \\
 Saudi Arabia &       2 \\
      Lebanon &       2 \\
\end{tabular}
\begin{tabular}{lc}
        Italy &       2 \\
     Slovakia &       2 \\
     Thailand &       1 \\
      Ireland &       1 \\
     Malaysia &       1 \\
     Scotland &       1 \\
       Taiwan &       1 \\
     Slovenia &       1 \\
   Bangladesh &       1 \\
    Argentina &       1 \\
     Colombia &       1 \\
      Croatia &       1 \\
  South Korea &       1 \\
        Korea &       1 \\
      Austria &       1 \\
  Switzerland &       1 \\
\end{tabular}
\caption{Total number of papers included in this study per country}
\label{tb:country}
\end{table}

\newpage
\section{Popularity per Tool Across Publications}
\label{ap:popularity}

\begin{table}[h]
\footnotesize
\begin{tabular}{lc}
\toprule
\textbf{Name} &  \textbf{Sum} \\\midrule
Scratch                  &  127 \\
Blockly                  &   59 \\
Alice                    &   42 \\
Snap!                    &   40 \\
App Inventor             &   40 \\
Lego Mindstorms          &   23 \\
PencilCode               &   13 \\
GreenFoot                &   12 \\
Makecode                 &   11 \\
GP                       &    9 \\
Tiled Grace              &    9 \\
Dr. Scratch              &    6 \\
Code.org                 &    6 \\
Scratch Jr               &    6 \\
Droplet                  &    5 \\
LogoBlocks               &    5 \\
PocketCode               &    5 \\
Ardublockly              &    5 \\
StarLogo TNG             &    5 \\
BYOB                     &    4 \\
NetsBlox                 &    4 \\
Nettango                 &    4 \\
Blocks4All               &    4 \\
Stencyl                  &    4 \\
CT-Blocks                &    4 \\
Blocklyduino             &    4 \\
BlockyTalky              &    4 \\
Hairball                 &    4 \\
RoboBuilder              &    3 \\
pencil.cc                &    3 \\
Deltatick                &    3 \\
DrawBridge               &    3 \\
Openblocks               &    3 \\
ArduBlock                &    3 \\
Modkit                   &    3 \\
Squeak eToys.            &    3 \\
Bags                     &    3 \\
Dash and dot             &    3 \\
Open roberta             &    3 \\
Hopscotch                &    3 \\
CoBlox                   &    2 \\
Storyblocks              &    2 \\
robot blockly            &    2 \\
BlockEditor              &    2 \\
SQLSnap                  &    2 \\
\end{tabular}
\begin{tabular}{lr}
BridgeTalk               &    2 \\
Tern                     &    2 \\
Ozobots                  &    2 \\
Waterbear                &    2 \\
Torino                   &    2 \\
Agentcubes               &    2 \\
mBlock                   &    2 \\
EvoBuild                 &    2 \\
UML                      &    2 \\
DataSnap                 &    2 \\
Ladder                   &    2 \\
Kodu                     &    2 \\
Blockly@rduino           &    2 \\
Tynker                   &    2 \\
Frog Pond                &    2 \\
iSnap                    &    2 \\
LaPlaya                  &    1 \\
ITCH                     &    1 \\
DBSnap                   &    1 \\
BEESM                    &    1 \\
CustomPrograms.1         &    1 \\
CodeIt                   &    1 \\
CARMEN                   &    1 \\
Pixly                    &    1 \\
BlockImpress             &    1 \\
Polymorphic Blocks       &    1 \\
ecraft2learn             &    1 \\
Spherly                  &    1 \\
FabCode                  &    1 \\
MakerArcade              &    1 \\
MORPHA                   &    1 \\
RoboBlockly              &    1 \\
edbot                    &    1 \\
NEPO                     &    1 \\
Patch                    &    1 \\
DStBlocks                &    1 \\
Neuroblock               &    1 \\
ROBOLAB                  &    1 \\
Amphibian                &    1 \\
GradeSnap                &    1 \\
Accessible Blockly       &    1 \\
PopBots                  &    1 \\
Quality Hound            &    1 \\
Prime                    &    1 \\
Pictogramming            &    1 \\
AliCe-VilLagE            &    1 \\
\end{tabular}
\begin{tabular}{lr}
TouchDevelop             &    1 \\
Calico Jigsaw            &    1 \\
BlueJ                    &    1 \\
SPARQL                   &    1 \\
Tickle                   &    1 \\
ScratchX                 &    1 \\
Flip                     &    1 \\
Logo                     &    1 \\
Code3                    &    1 \\
Snap4Arduino             &    1 \\
E-Block                  &    1 \\
RoboBlocks               &    1 \\
Microsoft Touch Develop  &    1 \\
MicroApp                 &    1 \\
Blockpy                  &    1 \\
AppLap                   &    1 \\
CustomPrograms           &    1 \\
Robobo                   &    1 \\
CodeSpells               &    1 \\
Turtle Art               &    1 \\
Micropython              &    1 \\
Sketchware               &    1 \\
Thunkable                &    1 \\
AppyBuilder              &    1 \\
BLOX                     &    1 \\
Finch robot              &    1 \\
App Inventor Java Bridge &    1 \\
Robokol                  &    1 \\
BehaviourComposer        &    1 \\
AgentSheets              &    1 \\
TurtleArt                &    1 \\
PicoBlocks               &    1 \\
Scratch Memories         &    1 \\
Scratch Community Blocks &    1 \\
PseudoBlocks             &    1 \\
OzoBlockly               &    1 \\
TinkerBlocks             &    1 \\
ViMAP                    &    1 \\
KidSim                   &    1 \\
Gameblox                 &    1 \\
Phratch                  &    1 \\
Romo                     &    1 \\
N-Bot                    &    1 \\
Cherps                   &    1 \\
Toon Talk                &    1 \\
StarLogoBlocksNova       &    1 \\
\end{tabular}
\begin{tabular}{lr}
Visual AgenTalk          &    1 \\
Tica                     &    1 \\
Blockly Turtle           &    0 \\
Cognimates               &    0 \\
\bottomrule
\end{tabular}
\end{table}

\newpage
\section{Programming languages used to implement \bbes}
\label{ap:pls}

\begin{table}[h]
	\centering
\footnotesize
\begin{tabular}{lc}
\toprule
\textbf{Programming language} &  \textbf{Papers} \\
\midrule
N/A                                        &                   100 \\
JavaScript                                 &                    15 \\
HTML, JavaScript, and CSS                  &                    15 \\
Java                                       &                     3 \\
Python                                     &                     2 \\
iOS (Swift)                                &                     2 \\
JavaScript \& Java                         &                     2 \\
Pharo Smalltalk                            &                     2 \\
TypeScript                                 &                     2 \\
AngularJs and TypeScript                   &                     1 \\
Snap!                                      &                     1 \\
Python, HTML, CSS and Javascript           &                     1 \\
Java / JastAdd                             &                     1 \\
Java / ANTLR                               &                     1 \\
ActionScript                               &                     1 \\
PHP-based framework \& JavaScript and HTML5 &                    1 \\
Tiled Grace                                &                     1 \\
\bottomrule
\end{tabular}
\caption{Programming languages used to implement \bbes.}
\label{tab:ap-pls}
\end{table}

%\newpage
\section{Tools used for development}
\label{ap:devtools}

\begin{table}[h]
	\centering
\footnotesize
\begin{tabular}{lc}
\toprule
\textbf{Library} & \textbf{\# of languages}  \\
\midrule
NaN                                &              62 \\
Blockly                            &              40 \\
Scratch                            &              11 \\
Snap!                              &               7 \\
Scratch 3.0 (Blockly)              &               3 \\
CT-Blocks                          &               3 \\
App inventor \& Blocky             &               2 \\
Microsoft MakeCode                 &               2 \\
BlocklyDuino                       &               2 \\
Python                             &               1 \\
NetTango                           &               1 \\
Blockly@rduino                     &               1 \\
Blockly and Standard pictogramming &               1 \\
Nettango                           &               1 \\
Blockly, Monaco, MakeCode          &               1 \\
iSnap~\cite{Price:2017}            &               1 \\
Droplet                            &               1 \\
MakeCode                           &               1 \\
Snap! \& Scratch                   &               1 \\
Deltatick                          &               1 \\
\end{tabular}
\begin{tabular}{lc}
NetsBlox                           &               1 \\
PencilCode, Droplet                &               1 \\
Openblocks~\cite{Roque:2007}       &               1 \\
Ardublock                          &               1 \\
Tiled Grace                        &               1 \\
Scratch \& Blockly                 &               1 \\
Snap! \& DB Snap                   &               1 \\
App inventor                       &               1 \\
\bottomrule
\end{tabular}
\caption{List of tools used for the development of \bb editors.}
\label{ap:meta-tools}
\end{table}

\newpage
\section{Domains}
\label{ap:domains}

\begingroup
\footnotesize
\begin{longtable}{lc}
\toprule
\textbf{Sub-category} & \textbf{Category} \\
\midrule
\endhead
%\midrule
%\multicolumn{2}{r}{{Continued on next page}} \\
%\midrule
\endfoot

\bottomrule
\endlastfoot
End-user programming & \multirow{7}{*}{Programming} \\
Parallel Programming &  \\
Behavioural programming &  \\
Audio programming &  \\
Collaborative Programming &  \\
Programming &  \\
Programming Models &  \\\midrule
Education & \multirow{11}{*}{Education} \\
Parallel Programming education &  \\
Computational Thinking &  \\
Programming education &  \\
socioscientific issues (SSI) education &  \\
Grammar education &  \\
Education programming &  \\
Computer Science Education &  \\
Simulation-based training &  \\
Latin &  \\
Parson problems &  \\\midrule
Programming environments & \multirow{6}{*}{Programming environments} \\
Novice programming environments &  \\
Accessible Programming Environments &  \\
Spreadsheets &  \\
Block-based Environments &  \\
Frame-based editing &  \\\midrule
Security & \multirow{2}{*}{Security} \\
Privacy &  \\\midrule
Robotics & \multirow{9}{*}{Physical computing} \\
Micro:bit &  \\
Collaborative robots &  \\
Embedded systems development &  \\
Physical computing &  \\
Microcontrollers &  \\
Automotive manufactoring &  \\
Smart cities &  \\
Mobile &  \\\midrule
Scratch & \multirow{3}{*}{Languages} \\
Visual DSLs &  \\
Block-based languages &  \\\midrule
Augmented reality & \multirow{11}{*}{HCI} \\
Human Robotics Interaction &  \\
Usability &  \\
Tangible surfaces &  \\
Intelligent Tutoring Systems &  \\
HCI &  \\
Brain-Computer Interface &  \\
Tangible user interfaces &  \\
Brain-computer Interfaces &  \\
end-user development &  \\
Accesibility &  \\\midrule
Visualization & \multirow{6}{*}{ARTS \& Creativity} \\
Gaming &  \\
Music &  \\
Cultural Heritage (museums \& art) &  \\
Tinkering &  \\
Storytelling &  \\\midrule
Biology & \multirow{3}{*}{Science} \\
Chemistry &  \\
Medicine &  \\\midrule
AI & \multirow{6}{*}{AI} \\
Data Science &  \\
Data-driven programs &  \\
Data Analysis &  \\
data analysis and visualization &  \\
Machine learning &  \\\midrule
program analysis & \multirow{7}{*}{Software Engineer} \\
Code quality &  \\
Testing &  \\
Databases &  \\
Static analysis &  \\
program analysis and transformation &  \\
Software quality &  \\\midrule
Agent-based computational model & \multirow{3}{*}{Agent-based model} \\
Agent-based modelling &  \\
Agent-based modelling tasks & \\
\end{longtable}
\endgroup

\end{document}